\documentclass[
]{ceurart}

\usepackage[utf8]{inputenc}  
\usepackage{inconsolata}  

\usepackage{amsmath, amsthm, amsfonts}

\usepackage{placeins}
\usepackage{graphicx}
\usepackage{caption}       
\usepackage{caption}
\usepackage{float}         
\usepackage{booktabs}      
\usepackage{multirow}      
\usepackage{array}         
\usepackage{tabularx}      
\usepackage{longtable}     
\usepackage{colortbl}      
\usepackage{arydshln}      

\usepackage{algorithm}
\usepackage{algorithmic}

\usepackage{listings}
\usepackage{newtxtt}
\usepackage{xcolor}  

\definecolor{codegreen}{rgb}{0,0.6,0}
\definecolor{codegray}{rgb}{0.5,0.5,0.5}
\definecolor{codepurple}{rgb}{0.58,0,0.82}
\definecolor{backcolour}{rgb}{0.95,0.95,0.92}

\usepackage{rotating}

\usepackage{hyperref}
\usepackage{url}

\usepackage{enumitem}

\usepackage{array}
\usepackage{breakurl}
\begin{document}

\copyrightyear{Copyright 2025 for this paper by its authors.}
\copyrightclause{Use permitted under Creative Commons License Attribution 4.0 International (CC BY 4.0).}

\conference{International Workshop on AI Governance: Alignment, Morality and Law (AIGOV) 2025. AAAI Conference on Artificial Intelligence.}

\title{Enhancing LLMs for Governance with Human Oversight: Evaluating and Aligning LLMs on Expert Classification of Climate Misinformation for Detecting False or Misleading Claims about Climate Change}


\author[1]{Mowafak Allaham}
\fnmark[1]
\address[1]{Northwestern University, 
  2240 Campus Drive Evanston, IL 60208, USA}

\author[2]{Ayse D. Lokmanoglu}
\address[2]{Boston University, 640 Commonwealth Avenue Boston, MA 02215, USA}

\author[3]{P. Sol Hart}
\address[3]{University of Michigan, 5417 North Quad 105 S. State St., Ann Arbor, MI 48109-1285, USA}

\author[1]{Erik C. Nisbet}

\fntext[1]{Corresponding author.}

\begin{abstract}
Climate misinformation is a problem that has the potential to be substantially aggravated by the development of Large Language Models (LLMs). In this study we evaluate the potential for LLMs to be part of the solution for mitigating online dis/misinformation rather than the problem. Employing a public expert annotated dataset and a curated sample of social media content we evaluate the performance of proprietary vs. open source LLMs on climate misinformation classification task, comparing them to existing climate-focused computer-assisted tools and expert assessments. Results show (1) open-source models substantially under-perform in classifying climate misinformation compared to proprietary models, (2) existing climate-focused computer-assisted tools leveraging expert-annotated datasets continue to outperform many of proprietary models, including GPT-4o, and (3) demonstrate the efficacy and generalizability of fine-tuning GPT-3.5-turbo on expert annotated dataset in classifying claims about climate change at the equivalency of climate change experts with over 20 years of experience in climate communication. These findings highlight 1) the importance of incorporating human-oversight, such as incorporating expert-annotated datasets in training LLMs, for governance tasks that require subject-matter expertise like classifying climate misinformation, and 2) the potential for LLMs in facilitating civil society organizations to engage in various governance tasks such as classifying false or misleading claims in domains beyond climate change such as politics and health science.
\end{abstract}

\begin{keywords}
  Human oversight \sep
  LLM alignment \sep
  Climate misinformation \sep
  AI governance.
\end{keywords}

\maketitle

\section{Introduction}
When considering public support for climate change science and policy, people rely on information from mediated sources such as online or offline media rather than directly from scientists \cite{scheufele2014science}. The reliance on such sources provide an opportunity for false or misleading claims about climate change to compete with accurate ones and influence public opinion and policy discourse which seriously hampers climate mitigation efforts \cite{allgaier2019science,gounaridis2024social,RN15,lewandowsky2021climate,treen2020online}. 

Advances in Large Language Models (LLMs) could also contribute to the exacerbation and diffusion of climate misinformation if exploited by malicious actors. LLMs are capable of generating high volumes of persuasive, deceptive, and human-like content full of misinformation that promotes climate change denial and skepticism \cite{fore2024unlearning,zhang2024toward,climate2024underperforming,ellison2024climate} making it more difficult for humans to detect LLM-generated dis/misinformation in news context especially when circulating on social media \cite{kreps2022all,marlow2020twitter}.

The major knowledge gap about the nature of climate change information on digital platforms, combined with inconsistent and ineffective content moderation policies across platforms \cite{climate2024underperforming,toxictenten2021,romero2023factsheet}, continue to seriously hamper the effective mitigation of climate change misinformation on social media. A key element of this problem is the inadequate identification and classification tools that capture the technical expertise required to evaluate the veracity of claims about climate change circulating online \cite{coan2021computer,vu2023fact}. 

In response, researchers have attempted to develop a variety of LLM-assisted classification tools to identify and respond to false or misleading claims about climate change \cite{coan2021computer,leippold2024automated}. However, the generalizability of these detectors is less robust on types and sources of text beyond what they are trained on, demonstrating how easy it is for malicious actors to overcome the detection by these algorithms \cite{stiff2022detecting}. Additionally, existing LLM benchmarks do not always represent real-word tasks \cite{ni2024mixeval} and typically frame climate misinformation detection as a binary problem \cite{thulke2024climategpt,lacombe2023climatex}. Existing evaluations of LLMs are predominantly centered around detecting the \textit{presence} of climate misinformation in text \cite{thulke2024climategpt}, rather than on identifying the \textit{types} of claims that are used to convey such misinformation. Thus, these benchmarks fall short from guiding domain and policy experts in selecting the most suitable model for detecting and classifying claims about climate change at a scale.

To address these issues, we first benchmark 
the performance of 11 open-source and 5 proprietary (i.e., closed-source) models, on a zero-shot task, in classifying false or misleading claims (FMC) about climate change using a public dataset on climate-misinformation \cite{coan2021computer}. This dataset, sourced from contrarian and skeptical domains about climate change, also includes the annotation of these claims by climate-communication experts, which serve as the ground truth for comparing LLM responses. Next, we selected the best performing LLM and took a human-centered approach into evaluating the generalizability of the LLM in detecting false or misleading claims about climate change present in articles sourced from low credible news sources that are circulating on social media. We incorporate the assessment of two experts with over 20 years of experience in climate communication into the evaluation of the capability of these LLMs in classifying claims that form the basis of climate misinformation. By pairing a quantitative evaluation with expert assessment of LLMs, we overcome some of the limitations related to the validity, quality, and diversity of automated benchmarks that are commonly used by the AI community when performing automated evaluation of LLMs \cite{gehrmann2023repairing,xiao2024human}. 

Within this context, we explore in this work four research questions:
\begin{enumerate}[label=\textbf{RQ\arabic*:}, left=0pt]
    \item How well do LLMs, both open-source and closed-source, classify false of misleading claims about climate change compared to classifications made by climate-experts? 
    \item How do these models compare to existing computer-assisted approaches? 
    \item Does fine-tuning on expert-annotated dataset enhance the model's ability to generalize better and classify false or misleading claims about climate change circulating on social media?
    \item How aligned are the classified claims by the fine-tuned model on social media data with domain expert annotations of these claims?
\end{enumerate}

The results of our study provide the following four contributions: (1) Although open-source LLMs perform well on evaluation benchmarks, we found that they substantially underperform compared to proprietary models in accurately classifying false or misleading claims about climate change on an expert-annotated dataset. (2) We observe and report an inferior performance of proprietary LLMs (GPT-4o, GPT-4o-mini, GPT-4, and Gemini-1.5-flash) as compared to a BERT-based computer assisted tool (CARDS) \cite{coan2021computer} for classifying false or misleading claims about climate change. (3) We demonstrate the superior performance of fine-tuning GPT-3.5-turbo, as compared to other proprietary models and CARDS \cite{coan2021computer}, for classifying false or misleading claims commonly found in climate skeptic and contrarian blogs as annotated by climate change experts. (4) We illustrate GPT-3.5-turbo's strong and extensible capability for classifying false or misleading claims about climate change commonly found in social media from sources beyond conservative think tanks and contrarian blogs with the approximate reliability as two senior climate change communication experts with over 20 years of experience. Through our findings, we also report a skewed distribution of the types of claims classified by the LLMs potentially reflecting an underlying bias in their training data.

We  hope for this research to contribute to the collective efforts aimed at assessing the efficacy of LLMs in detecting and guardrailing against climate dis/misinformation. In addition, we encourage AI researchers, designers, and developers to reflect on our approach and findings as they consider deploying LLMs for content moderation or as part of automated evaluation workflows for identifying false or misleading claims about climate change in either human or AI generated content.

\section{Related Work}
The potential abuse of LLMs by malign actors to generate misinformation that shapes the information environment and public opinion has become evident across a range of domains from politics to healthcare, and including climate change \cite{Yang_2024,marlow2020twitter,ferrara2020characterizing,akhtar2023false,de2023chatgpt}. However, detecting climate dis/misinformation generated by humans is difficult, let alone content generated by LLMs \cite{chen2023can}. In an effort to combat climate dis/misinformation, researchers have introduced tools that leverage LLMs to identify false claims about climate change from skeptic and contrarian sources, fact-check these claims, and even generate factual and in-depth answers to climate related questions \cite{coan2021computer,leippold2024automated,thulke2024climategpt,mullappilly2023arabic}. Despite these efforts, researchers have questioned the generalizability of dis/misinformation detection tools and have identified vulnerabilities that enable malicious content to bypass them \cite{stiff2022detecting}. One proposed solution to this issue is to establish benchmarks to evaluate the capability of LLMs in classifying and detecting climate related content.

Researchers have constructed several datasets relevant to climate change research, but not always tailored for benchmarking the detection of specific types of climate dis/misinformation. Existing datasets are being used to evaluate LLMs for climate-related classification tasks such as: predicting sentence relevance to climate change \cite{leippold2024automated}, stance detection in support or opposition toward climate change prevention \cite{vaid-etal-2022-towards}, and fact-checking scientific information related to climate science \cite{laud2023climabench,pirozelli2023benchmarks}.
    
Still, such benchmarks seem to be suffering from limitations in terms of the diversity \cite{ni2024mixeval}, complexity \cite{liang2022holistic}, and representation of how climate dis/misinformation is being manifested in terms of false or misleading claims \cite{thulke2024climategpt}. It is essential, therefore, to involve stakeholders such as climate change experts in the design, development, and validation for many of these tools and datasets, as their input and feedback not only refines the quality of the benchmark datasets, but also contributes to model enhancements in terms of handling climate misinformation \cite{stiennon2020learning,zhou2020learning,ouyang2022training,christiano2017deep}. Through such human-centered approach, stakeholders can identify in-depth criteria for what these models are being evaluated on and propose edge cases that impose constraints on the model performance before it gets deployed or leveraged for governance tasks 
 \cite{xiao2024human}.

\section{Data}
Our study employs two datasets: 1) a public dataset of false or misleading claims about climate change used to train Climate Change Denial and Skepticism (CARDS) model \cite{coan2021computer} and 2) a curated sample of the most engaging articles and blog posts (i.e., based on the number of likes, comments, shares) about climate change on Facebook and X (i.e., formerly known as Twitter) from right-biased, questionable, and low credible sources.

\subsection{Evaluating LLMs using CARDS}\label{3.1}
The curated dataset to train the CARDS model contains paragraphs in English language from articles sourced from 53 contrarian and skeptical domains about climate change spanning the years 1998 to 2020 and their corresponding claims based on annotations from the authors who are experts in climate research \cite{coan2021computer}. 
The dataset is randomly split into training (N=23,436), validation (N=2,605), and test sets (N=2,904). Our decision to use this dataset for evaluating LLMs is based on the (1) breadth of this dataset on climate misinformation and (2) the taxonomy of claims developed and validated by experts in climate-research \cite{coan2021computer}. In addition, the availability of the taxonomy of super-claims and coding manuals used to annotate the claims also enables us to craft prompts using the instructions in these manuals as part of the zero-shot task to classify false and misleading claims about climate change (as later described in section \ref{4.1}).
The CARDS taxonomy of super claims and sub-claims represents the primary arguments employed by climate denialists and skeptics (see Appendix \ref{a.1}). False and misleading claims in this taxonomy are grouped into five main categories: (1) global warming is not happening, (2) humans greenhouse gases are not causing global warming, (3) climate impacts are not bad, (4) climate solutions won’t work, and (5) the climate movement and/or science are unreliable. Claims in the CARDS dataset are formatted as strings that combine the super-claim and sub-claim into a single label separated by an underscore (e.g., "5\_1") referring to the super-claim and sub-claim, respectively.

\subsection{Social media dataset}\label{3.2}
\textbf{Scraping data.} Using an API from NewsWhip, a social media analytics platform, we retrieved the URLs for the daily 5000 most engaging English-language articles discussing climate change, per their accumulated number of likes, shares, and comments, that were published from American domains on Facebook and X between January and December 2022, inclusive. A list of relevant keywords compiled by a climate change communication expert was used to query the NewsWhip API. A full list of the keywords can be found in the Appendix (see Appendix \ref{a.2}). Next, we scraped the text and metadata (e.g., publication date) from all the URLs retrieved from NewsWhip using a custom web scraper in Python that leverages Newspaper3K and BeautifulSoup libraries.
    
We ensured the scraped articles were centered on climate change, and not merely mentioning the topic in passing, by filtering the corpus using the same list of keywords in Appendix \ref{a.2} used for the API search. The keyword filter was applied to the headline and first 250 words of each scraped article, resulting in a corpus of 829,827 articles with climate change and published on Facebook or X in 2022.

\textbf{Domain credibility}. We then filtered our corpus by the domain credibility of the publisher as means to curate a test dataset that contained a sufficient number of false or misleading claims about climate change circulating on social media. 
To determine the domain level credibility of each scraped article, we relied on a combination of the Media Bias Fact Check (MBFC) categories \cite{mbfc} and NewsGuard Trust score \cite{newsguard}. MBFC categorizes news sources in one of nine bias categories: least biased, left bias, left-center bias, right-center bias, right bias, conspiracy-pseudoscience, questionable sources, pro-science, and satire. Similarly, NewsGuard Trust score is a reliability rating between 0 and 100 that is assigned by journalists and editors to news websites based on journalistic and apolitical criteria such as credibility and transparency. A NewsGuard score of 60 or below indicate an untrustworthy news source.
    
Employing these two resources, we appended the MBFC category and the NewsGuard score to all articles in our corpus with domains matching those in the two datasets. We then selected all articles from MBFC right-bias, conspiracy-pseudoscience, and questionable categories and/or had a Newsguard score of 60 or below as articles that are most likely to contain false or misleading information. Out of the 829,927 articles published about climate change in 2022 on Facebook and X, 71,175 (8.6\%) were classified as originating from right-bias, conspiracy-pseudoscience, and questionable domains with low credibility\footnote{Check Appendix \ref{a.3} for details about the most prevalent domains in our dataset}.

\section{Methodology}
\subsection{Benchmarking LLMs using CARDS}\label{4.1}
To assess how well open-source vs. proprietary LLMs perform in classifying false or misleading claims about climate change, we apply zero-shot classification technique to classify paragraphs about climate change, sourced from articles published by conservative think tanks and blog posts, that are in the CARDS test dataset. This dataset serves as a baseline for comparing the claims classified by each LLM with the annotated claims by climate-research experts in the CARDS test dataset on the paragraph level. All evaluations of the 11 open-source models in this research were conducted on Google Colab, which constrained our model selection process for this task based on the availability of the compute and memory resources. A full list of the LLMs used for this assessment and their performance can be found in Table \ref{tab:model-comparison-on-cards-test}. Open-source models in the aforementioned list were selected based on the limited computing resources we had access to on Google Colab. In contrast, for closed-source models, we rely on OpenAI models and Gemini because these models are widely used in various applications, including climate research \cite{zhu2023climate,kraus2023enhancing,mullappilly2023arabic}. This limitation reflects a similar constraint faced by civil organizations interested in this research and application area but lacking the resources to host high-performing open-source models. Along with base models, we chose to included instruction-tuned models because they are generally aligned to mitigate social harms, such as misinformation and manipulation, and tend to demonstrate improved reasoning \cite{achiam2023gpt,dubey2024llama}, which may translate to a better performance on our benchmarking task.


To prompt the different LLMs to classify false or misleading claims about climate change, we crafted two prompts: a prompt that reflects the coding manual of CARDS (\ref{a.4}) , and a more synthesized prompt (\ref{a.8}) that includes the same categories of claims in the CARDS model so the prompt fits the limited context window for some open source models.

For closed source models, we included a system and user prompts (\ref{a.4} and \ref{a.5}, respectively) based on the instructions derived from the coding manual used to train annotators for labeling the training and testing datasets that were subsequently used to train the RoBERTa$_{\mathit{large}}$ CARDS model \cite{coan2021computer}.

Similarly, we structured the prompts for the open-source models to reflect a question-answer prompt template as illustrated in \ref{a.8}. This structure is commonly used for evaluating LLMs in zero-shot tasks \cite{rajpurkar2018know}. 

The phrasing of the question in the prompts for both open-source and proprietary models was derived from the CARDS coding manual. We crafted these prompts to be consistent with the claims definitions in the CARDS coding manual so we have a one-to-one comparison of the performance between the classified claims by the LLMs and the annotated claims by expert coders. Next, we validated the prompts on the CARDS validation dataset to confirm that the models are compliant with their corresponding prompts and that the generated claims are in the same format as in the CARDs datasets.

To evaluate the performance of open-source and proprietary LLMs in classifying claims about climate change, we leveraged the test set (N=2,904) described in section \ref{3.1}.

\textbf{Classifying claims using LLMs}. To classify claims using proprietary models, we leveraged ``openai'' and Google's ``genai'' libraries to classify each paragraph in the test dataset by sending requests to OpenAI GPT models (GPT-4o, GPT-4o-mini, GPT4, and GPT-3.5-turbo) and Gemini-1.5-flash API endpoints, respectively. Each request includes the system and user prompts outlined in Appendix \ref{a.4} and \ref{a.5}. As for opens-source models, we employed the transformers library to load each open-source model in 8-bit quantization from HuggingFace and used prompt \ref{a.8} to instruct each model to classify paragraphs for false or mis-leading claims.

To ensure the reproducability of the labels for the classified claims by each model and avoid having the models generate a preamble messages as part of the response, we configured the temperature at inference to 0 and 0.001 \footnote{Huggingface TGI inference endpoint doesn't allow setting the temperature to 0.} for proprietary and open-source models, respectively.

Although we prompted the models at the sub-claim level to match CARDS training dataset's conceptual framework structured around specific subclaims, as outlined in the original CARDS paper \cite{coan2021computer}. However, for analysis and applications aimed at refuting climate misinformation, we will focus only at the super-claim level. Accordingly, to assess the performance of the different models in classifying super-claims, we compare the classified claims using the LLMs with the annotated claims in the CARDS test dataset at the super-claim level of analysis. To this end, we split the LLM-generated claim labels and those in the test dataset based on the under score delimiter. Only the number to the left of the underscore ("\_") delimiter was extracted for model evaluation and comparison as it represents the super-claim. We elaborate on our findings from benchmarking the performance of the different LLMs with respect to the test dataset of the CARDS model in section \ref {5.1} of the results section. 


\subsection{Fine-tuning GPT-3.5-turbo on CARDS}\label{4.2}
The need to constantly update, test, and validate prompts, especially for emerging claims, makes zero-shot prompting a less scalable approach when compared to fine-tuning. Fine-tuning enhances the ability of general purpose LLMs to be more optimal in their adaptability to domain-specific tasks such as identifying emerging types of claims about climate change by including a few examples of these claims in the training data. In addition, fine-tuned LLMs that accurately classify false or misleading information about climate change have the potential to be integrated into a variety of governance tools such as those used for content moderation \cite{leippold2024automated}.


For these reasons, along with our findings from benchmarking different LLMs as elaborated in section \ref{5.1} of results, we chose to fine-tune GPT-3.5-turbo\footnote{Fine-tuning GPT-4o was not yet available for fine-tuning at the time of conducting this analysis.} to classify and compare the classified false or misleading claims about climate change of the fine-tuned model with the claims classified by the other models in the zero-shot task described in section \ref{4.1}, and those annotated in the CARDS test set by climate-research experts. 

Employing OpenAI's Python library, we fine-tune GPT-3.5-turbo for 3 epochs with a batch size of 8 using the training (N=23,436) and validation (N=2,605) datasets from the CARDS. The number of epochs and batch size were selected to closely resemble the training hyperparamters for RoBERTa$_{\mathit{large}}$ CARDS model\footnote{RoBERTa$_{\mathit{large}}$CARDS model was trained on 3 epochs and batch size of 6.}. 

For the purpose of fine-tuning, we structured the CARDS training and validation datasets in a list of chat message dictionaries as recommended by OpenAI\footnote{\url{https://platform.openai.com/docs/guides/fine-tuning/example-format}}. Each dictionary includes a system and user messages as shown in Appendix \ref{a.6}. All compiled dictionaries of system and user messages were grouped in a JSONL file format to comply with OpenAI fine-tuning requirements. Accordingly, GPT-3.5-turbo was fine-tuned on 13,330,863 tokens for 8,789 steps and costed \$106.65 USD. 

To assess the performance of the fine-tuned model with respect to the RoBERTa$_{\mathit{large}}$ CARDS model, we used the fine-tuned model to classify each paragraph in the CARDS test set into its corresponding claim label. We leveraged OpenAI's Python library to send  formatted requests as chat messages, similar to what we have already done for zero-shot classification in section \ref{4.1}, to the model's chat completion endpoint. We elaborate on the findings from our assessment in section \ref{5.1} of the results. 

Next, we describe how the fine-tuned GPT-3.5-turbo and the RoBERTa$_{\mathit{large}}$ CARDS models were applied to classify false or misleading claims about climate change in paragraphs from most engaging and low-credibility articles on social media about climate change.

\subsection{Classifying claims in  social media data}\label{4.3}
Employing the social media dataset described in section \ref{3.2}, we utilize the fine-tuned GPT-3.5-turbo to classify claims underpinning misinformation about climate change at the paragraph level similar to CARDS. The fine-tuned model codes each paragraph in each article in this dataset as containing either (1) no false or misleading claim or (2) one of 27 false or misleading claim labels outlined in the taxonomy of claims of the CARDS model in Appendix \ref{a.1}. The claim labels combine the super-claim and sub-claim into a single label, delimited by an underscore ("\_"). An example of the prompt structure used for inference is illustrated in Appendix \ref{a.7}. 

A total of 856,722 paragraphs from 71,175 articles published by low credible domains between January and December 2022 were classified by the fine-tuned model. We randomly selected a stratified sample of 914 paragraphs and their corresponding claim labels for manual evaluation by a pair of climate change communication experts. Half of the paragraphs were selected randomly from the generated claims that were labeled to have no claim and the other half was randomly selected using stratified sampling from all other types of claims in proportion to their distribution within the 2022 social media dataset. 

We also apply RoBERTa$_{\mathit{large}}$ CARDS model on the sampled paragraphs from social media by retrieving the model weights from Github\footnote{\url{https://github.com/traviscoan/cards}} and classifying the claims in these paragraphs to compare the performance of the fine-tuned model to the computer-assisted CARDS approach for classifying false or misleading claims about climate change. 

After classifying the claims in the sampled paragraphs from most engaging articles on social media using the fine-tuned GPT-3.5-turbo and RoBERTa$_{\mathit{large}}$ CARDS models, two climate communication experts manually coded for the claims in the same set of paragraphs. This provides a baseline to compare the efficacy of the aforementioned LLMs with respect to expert classification of climate misinformation. We elaborate on the coding procedure in the next section.

\subsection{Expert annotation of claims in social media data}\label{4.4}
Two climate change communication experts, each with over 20 years of experience as professors and academic researchers on how climate science is communicated, annotated the sampled paragraphs from mostly engaging articles on social media (described in section \ref{4.3}) on the super-claim and sub-claim levels per the CARDS coding manual\cite{coan2021computer}.

Each annotator separately reviewed each paragraphs and assigned a corresponding super-and sub-claims based on its content. Coding the claims on these two levels enable the annotators to detect claims at a more granular level, which allows for the identification and validation of super-claims based on the detected sub-claims.
The annotators then consulted and resolved any disagreements to create a final reconciled dataset of expert labels for all 914 paragraphs. The resulting labels from the manual annotation process then were formatted to include the super-claim and sub-claim in a single label that resembles the formatting used to train CARDS and fine-tune the GPT-3.5-turbo model\footnote{The formatting is similar to the one described in section \ref{4.1} combining the super- and sub-claims into a single label delimited by "\_"}.

The resulting labels from annotating the sampled paragraphs by the climate communication experts established a baseline (i.e., ground truth) to evaluate the level of alignment between the claims classified by GPT-3.5-turbo and RoBERTa$_{\mathit{large}}$, and the expert annotation of these claims on the social media data as described in section \ref{5.2}.

\section{Results}
In this section, we (1) compare the performance of open-source and proprietary models to the CARDS model on the CARDS test set to establish a performance baseline for classifying false or misleading claims, and (2) describe the level of alignment between the classified claims by the fine-tuned LLM and the expert annotations on the sampled paragraphs from the most engaging articles on social media that are published by low credible sources.

\subsection{Evaluating the performance of LLMs on CARDS data}\label{5.1}
Using the zero-shot approach (described in section \ref{4.1}) to classify climate misinformation on CARDS test dataset, we encountered some inconsistencies in the way some LLMs are responding to our task. Figure \ref{fig:taxonomy-of-claims} in the appendix shows the rate of valid responses per model. The prominent source of this inconsistency is the failure of models to comply with the instructions outlined in the prompt, resulting in responses that are irrelevant or with unstructured formatting. A similar inconsistent behavior of LLMs was also reported in prior research evaluating open-source LLMs on 20 NLP tasks \cite{bavaresco2024llms}. However, we observed instances of Llama3.1-8B failing to classify claims because it is "not sure if this is the right place to post this [request]" and that the model is "trying to get a better".

\begin{table}[h]
\centering
\small
\begin{tabular}{lccc}
\toprule
\textbf{Model} & \textbf{Precision} & \textbf{Recall} & \textbf{F1-Score} \\
\midrule
GPT-3.5-turbo (fine-tuned)& 0.88 & 0.81 & 0.84 \\
RoBERTa$_{\mathit{large}}$ (CARDS) & 0.82 & 0.75 & 0.77 \\
GPT-4o & 0.70 & 0.84 & 0.75 \\
GPT-4o-mini & 0.70 & 0.84 & 0.75 \\
GPT-4 & 0.70 & 0.82 & 0.74 \\
Gemini-1.5-flash & 0.63 & 0.79 & 0.67 \\
GPT-3.5-turbo (zero-shot) & 0.50 & 0.70 & 0.53 \\
Mistral-7B-Instruct-v0.3 & 0.36 & 0.52 & 0.28 \\
Meta-Llama-3.1-Instruct-8b & 0.28 & 0.41 & 0.25 \\
Phi-3.5-mini-instruct & 0.50 & 0.31 & 0.23 \\
Meta-Llama-3-8B-Instruct & 0.25 & 0.30 & 0.17 \\
Gemma-2-9b & 0.28 & 0.27 & 0.17 \\
Llama-2-7b-hf & 0.16 & 0.16 & 0.11 \\
Meta-Llama3.1-8b & 0.18 & 0.20 & 0.10 \\
Gemma-2-2b & 0.17 & 0.17 & 0.09 \\
Llama-2-7b-chat-hf & 0.20 & 0.18 & 0.06 \\
Meta-Llama-3-8B & 0.27 & 0.17 & 0.04 \\
Mistral-7B-v0.3 & 0.24 & 0.17 & 0.04 \\
\bottomrule
\end{tabular}
\caption{Performance results from Benchmarking open-source and proprietary LLMs on CARDS test dataset (N=2,904). All performance metrics are macro-averaged across the five categories of super-claims.}
\label{tab:model-comparison-on-cards-test}
\end{table}

To ensure a valid performance comparison across models, we replaced invalid responses from LLMs with values randomly sampled from a total of 28 possible labels that include the no claim label and 27 claim labels representing the super- and sub-claims categories of the five super-claims we are focusing our study on as outlined in the inference prompts (see \ref{a.7} and \ref{a.8}) and the claims taxonomy from CARDS illustrated in \ref{a.2}. 

By evaluating the models only at the super-claim level of analysis (as described in \ref{4.1}), our findings show a substantial gap in performance between open-source and proprietary models in classifying claims in paragraphs from the CARDS test dataset. Among the proprietary models, GPT-4o (F1$_{\text{macro}}$=0.75) and GPT-4 (F1$_{\text{macro}}$=0.74) outperformed Gemini-1.5-flash (F1$_{\text{macro}}$=0.67) on the zero-shot task across the five super-claims. In contrast, the highest scoring open-source model, Mistral-7B-Instruct-v0.3 achieved a macro-averaged F1-score of 0.28 over the five categories of super-claims, which is much lower than the performance of GPT-4o and GPT-4, as shown in Table \ref{tab:model-comparison-on-cards-test}.

We speculate that the performance gap between the models could be attributed to several factors, including the selection criteria of the data used for pre-training and instruction-tuning, as well as the decisions made as part of red-teaming these models \cite{achiam2023gpt,dubey2024llama}. For instance, despite deriving 50\% of the  tokens in its pre-training from general knowledge on the web, Llama-3.1-8B performed poorly (F1$_{\text{macro}}$=0.10) in classifying false or misleading claims about climate change \cite{dubey2024llama}. In addition, Meta's safety standards appear to be more focused on filtering out personal information from web content and preventing the the model from generating harmful content in any of the 13 hazard categories identified by Videgen et al.\cite{vidgen2024introducing}, which do not include a category for climate misinformation, may have impaired the model's ability to accurately identify and classify false information about climate change. In contrast, we observe GPT-4o's (F1$_{\text{macro}}$=0.75) performance to be substantially better than both completion (F1$_{\text{macro}}$=0.10) and instruction-tuned (F1$_{\text{macro}}$=0.25) Llama3.1-8B models. This could be due to OpenAI's initiative to red-team GPT-4o through partnering with over 70 experts in several domains, including experts on misinformation\footnote{https://openai.com/index/hello-gpt-4o/}. This collaborative effort appears to also have positively improved the model performance at least with respect to our classification task, as reflected by the uplift in performance between GPT-3.5 (F1$_{\text{macro}}$=0.53) and GPT-4o (F1$_{\text{macro}}$=0.75).

In comparison to the CARDS model (F1$_{\text{macro}}$=0.77), we found a comparable performance of GPT-4 (F1$_{\text{macro}}$=0.74) and GPT-4o (F1$_{\text{macro}}$=0.75) in classifying false or misleading claims about climate change as shown in Table \ref{tab:model-comparison-on-cards-test}. However, in comparison to CARDS, GPT-4o and GPT-4 appear to have a higher rate of false positives (Precision=0.70) indicating that the model is being more conservative in classifying paragraphs with no claims as false or misleading claims about climate change compared to the CARDS models. In addition, GPT-4o had the highest recall of all models used for zero-shot classification compared to CARDS indicating the model's capability to correctly classify instances containing claims relevant to climate misinformation.

Though GPT-4o is a much larger model compared to RoBERTa$_{\mathit{large}}$, in terms of parameter size, it is more budget friendly for stakeholders in climate change discourse such as researchers, policymakers, or think tanks to utilize this model when classifying climate misinformation as it does not require substantial investment in computational resources or expertise to develop a customized model such as CARDS for classifying climate misinformation. However, detecting false or misleading claims does not include the models' capability to respond to or mitigate such claims unless they are augmented with relevant external knowledge as seen in use cases for fact-checking climate change claims\cite{leippold2024automated}. Accordingly, future research is needed to explore and extend our work to determine whether models capable of detecting climate misinformation can augment general models, such as GPT-4o, with information about which claims are false or misleading to help orient these general models toward guard-railing against such claims.

Although GPT-4o and GPT-4 had comparable performance to CARDS in classifying false or misleading claims on the CARDS test dataset, fine-tuning GPT-3.5-turbo resulted in an even more performant model. The fine-tuned model (as described in section \ref{4.2}) has outperformed GPT-4o and RoBERTa$_{\mathit{large}}$ models on F1$_{\text{macro}}$ by 12.0\% and 9.1\%, respectively (see Table \ref{tab:comparison}). Furthermore, the fine-tuned model has an uplift in precision compared to GPT-4o and RoBERTa$_{\mathit{large}}$ CARDS model by 25.7\% and 7.3\%, respectively. This improvement was at a cost of a slight decline in the recall of the fine-tuned model (81\%) from GPT-4o's 84\%. Still, the fine-tuned model had a much better recall (81\%) compared to  RoBERTa$_{\mathit{large}}$ as shown in Table \ref{tab:comparison}.
In the next section, we describe our findings from evaluating the alignment between the claims classified using the fine-tuned GPT-3.5-turbo and RoBERTa$_{\mathit{large}}$), and those annotated by two climate-communication experts, each with over 20 years of experience in climate misinformation, on sample of paragraphs sourced from articles circulating on social media that are published by low credible sources.

\subsection{LLM vs. Expert classification of claims}\label{5.2}
First we evaluated the intercoder-relability to ensure the alignment between the two experts by calculating Krippendorf's alpha. Coders scored $_{\alpha_{\text{Krippendorff}}}$=.89 showing strong alignment between the two experts with respect to their coding of claims of the sampled paragraphs from articles about climate change circulating on social media. Then, using the claims classified by GPT-3.5-turbo and RoBERTa$_{\mathit{large}}$ we calculated the $_{\alpha_{\text{Krippendorff}}}$ for each model with respect to expert annotations.
    
We found a much higher level of alignment ($\displaystyle _{\alpha_{\text{Krippendorff}}}$=0.89) between the fine-tuned GPT-3.5-turbo and the climate research experts compared to RoBERTa$_{\mathit{large}}$ CARDS model ($_{\alpha_{\text{Krippendorff}}}$=0.66). This suggests that the fine-tuned GPT-3.5-turbo can classify false or misleading claims about climate change with the approximate reliability as two senior climate change communication experts with over 20 years of experience (see \ref{model-expert-comparison-on-sm-data} for examples). This opens up an opportunity for designing future AI systems that code and annotate climate misinformation at a scale with a human oversight.

\begin{table*}[ht]
\small
\centering
\begin{tabularx}{\textwidth}{c>{\centering\arraybackslash}X>{\centering\arraybackslash}X>{\centering\arraybackslash}X>{\centering\arraybackslash}Xc>{\centering\arraybackslash}X>{\centering\arraybackslash}X>{\centering\arraybackslash}X>{\centering\arraybackslash}X}
\hline
 & \multicolumn{4}{c}{\textbf{Fine-tuned GPT-3.5-turbo}} & & \multicolumn{4}{c}{\textbf{RoBERTa$_{\mathit{large}}$}} \\
\cline{2-5} \cline{7-10}
 \textbf{Claim Code} & \textbf{Precision} & \textbf{Recall} & \textbf{F1} & \textbf{Support} & & \textbf{Precision} & \textbf{Recall} & \textbf{F1} & \textbf{Support} \\
\hline
0 & 0.94 & 0.94 & 0.94 & 491 & & 0.76 & 0.96 & 0.85 & 491 \\
1 & 0.71 & 1.00 & 0.83 & 24 & & 0.71 & 0.62 & 0.67 & 24 \\
2 & 0.93 & 0.93 & 0.93 & 14 & & 1.00 & 0.79 & 0.88 & 14 \\
3 & 0.57 & 1.00 & 0.73 & 8 & & 0.57 & 0.50 & 0.53 & 8 \\
4 & 0.95 & 0.94 & 0.95 & 274 & & 0.93 & 0.63 & 0.75 & 274 \\
5 & 0.99 & 0.89 & 0.94 & 103 & & 0.90 & 0.62 & 0.74 & 103 \\
\hline
\textbf{Accuracy} &  &  & 0.94 & 914 & &  &  & 0.81 & 914 \\
\textbf{Macro avg} & 0.85 & 0.95 & 0.88 & 914 & & 0.81 & 0.69 & 0.74 & 914 \\
\textbf{Weighted avg} & 0.94 & 0.94 & 0.94 & 914 & & 0.83 & 0.81 & 0.80 & 914 \\
\hline
\end{tabularx}
\caption{Model performance comparison between the fine-tuned GPT-3.5-turbo and CARDS model in classifying false or misleading claims about climate change on a sample content from social media. Performance is measure assessed based on precision, recall, F1 scores. The claim labels corresponding to the claim codes are: (0) No claim, (1) global warming is not happening, (2) humans greenhouse gases are not causing global warming, (3) climate impacts are not bad, (4) climate solutions won’t work, and (5) the climate movement and/or science are unreliable. }
\label{tab:table2}
\end{table*}

Delving further into the performance comparison between GPT-3.5-turbo and the RoBERTa$_{\mathit{large}}$ CARDS model with respect to expert annotations at the super-claim level of analysis, we find the macro averaged F1 score for the fine-tuned GPT-3.5-turbo (F1$_{\text{macro}}$=0.88) to be 18.9\% higher than the one reported by RoBERTa$_{\mathit{large}}$ CARDS model (F1$_{\text{macro}}$=0.74) as shown in Table \ref{tab:table2}. The fine-tuned GPT-3.5-turbo also predominately had higher F1 scores across the five main categories of claims indicating strong performance by the model in identifying and classifying the main categories of claims outlined by the CARDS taxonomy, but on a broader sample of text from social media\footnote{\url{https://github.com/nwccpp/climatechange/tree/main/sample_data/aigov-expert-coded-sample}}.
        
On the other hand, we observed a poor performance by the fine-tuned model in detecting claims related to climate impacts are not bad. Reviewing the annotated super-claims and sub-claims by experts, we found that the fine-tuned model is unable to accurately classify sub-claims within this category about the impacts of climate change on animal and plant species (see sub-claim 3.2 within the taxonomy of claims in Appendix \ref{a.1}). We found that the fine-tuned GPT-3.5-turbo is biased toward inaccurately classifying  claims regarding climate change impacts on animal and plant species as false, possibly due to biases in the CARDS training data that we originally fine-tuned the model on (see Limitations section). This indicates additional fine-tuning is needed to enhance the ability of the model to differentiate between text describing positive versus negative impacts of climate change.

\section{Conclusion}
As developers and researchers test the potential of LLMs to persuade and misinform at scale \cite{matz2024potential,zhang2024toward}, evaluating the potential for LLMs to be part of the solution for governing online dis/misinformation rather than the problem becomes a task of great importance. In this context, the overarching goal of this paper was to demonstrate the crucial role of human-oversight when leveraging LLMs for governance by evaluating the performance of, both open-source and proprietary, LLMs in the application of classifying false or misleading claims about climate change. We also aimed to compare (1) how these models performed against existing tools designed for the same task, and (2) with the assessments made by climate-communication and mis-information experts.

The results showed that proprietary models have outperformed open-source LLMs in a zero-shot classification of climate change misinformation. Out of the proprietary models, and despite its inferior performance in the zero-shot task, we demonstrated how fine-tuning GPT-3.5-turbo model was superior to a trained BERT-based model and functionally equivalent to GPT-4o and a climate change communication expert with 20 years of experience in classifying claims about climate change in social media.

Though open-source LLMs performed poorly compared to proprietary models, they remain to have the potential for wide adoption by civil society organizations to engage in a range of important governance tasks (e.g., identifying and tracking dis/misinformation and hate speech). Accordingly, it is recommended for entities developing these models to (1) enhance the accessibility and transparency of the data used to train open-source models, and (2) start incorporating expert-opinions as part of the model performance feedback loop so the models can be leveraged, with human oversight, in tasks that may require experts such as classifying false or misleading claims in domains beyond climate change such as politics and health science.

\section*{Limitations}\label{limitations}
There are several limitations of this research. First, the data used for benchmarking GPT-4 and fine-tuning GPT-3.5-turbo is in English and in text format, which excludes claims in other languages and modalities (e.g., images and videos). This sets an important boundary condition on the performance of the models in identifying climate change misinformation while providing pathways for future research on LLM capabilities to accurately and reliably classify such content. 

Another limitation is the reliance of our research on a single taxonomy of claims developed by the authors of CARDS \cite{coan2021computer} that was leveraged to fine-tune GPT-3.5.-turbo. This introduces two sets of  biases. First, as the CARDS annotated dataset was based on an expert review of climate skeptic and contrarian domains, whether the GPT-3.5-turbo model is capable of precisely discriminating between accurate and inaccurate climate change claims within high credible sources (e.g. The New York Times, CNN, etc.) is an open question. One way to address this problem, as the authors of the CARDS model have recently moved toward, is a two-stage approach for classifying claims that first determines the veracity of the claim and in the second stage labels the category of false claims \cite{rojas2024augmented}. A next step, therefore, is to benchmark GPT-3.5-turbo model's performance in classifying claims from high credible sources for comparison to the updated CARDS model and continue fine-tuning as necessary on text sourced from domains with varying credibility.

A second model bias is the inability to classify claims in social media posts that do not fall within the CARDS taxonomy. For instance, the original taxonomy upon which CARDS is based includes claims about human health impacts as a category but was excluded in the final model due it its low prevalence \cite{coan2021computer}. However, this may be a function of the ideological-skew of climate skeptic blogs on which CARDS was trained that ignored this dimension of climate change impacts and/or temporal trends increasing the prominence of health impacts. As a result, climate change communication experts annotating the social media claims observed a substantial number of false claims about the health impacts of climate change on humans that did not fall within the current CARDS model and which the GPT 3.5-Turbo was unable to classify.

In addition, the model's poor performance in classifying claims about the impacts of climate change on animal and plant species (see Table \ref{tab:table2}), could also be attributed to the under-representation of these examples in the CARDS dataset used for fine-tuning. Moving forward, fine-tuning the GPT-3.5-turbo model on additional expert annotated datasets, for example from Climate Feedback \footnote{\url{https://science.feedback.org}} would likely enhance the model's performance in accurately classifying a wider range of claims. These limitations stress the importance of benchmarking the performance of LLMs against data collected from ``the wild'' as we did in this paper and fine-tuning accordingly to ensure optimal performance in detecting misinformation online. 

\section*{Ethics Statement}
Incorporating the knowledge of domain experts into the design and development of AI tools for classifying the veracity of claims about climate change, or any other topic (e.g., politics, healthcare, or public policy), requires careful considerations of bias and impact. Frameworks or taxonomies of “truth” integrated with computer-assisted tools, regardless of their scientific basis, may have ideological or inadvertent subjective biases that narrow the range of information that is deemed accurate or inaccurate beyond what is optimal for free and open discourse. Therefore, it is important to mitigate such biases in the design and development process of AI-driven claim-detection and fact-checking tools by incorporating the inputs from diverse teams of researchers.

The deployment of AI tools, similar to the ones evaluated in this work, to detect false or misleading claims also have social implications. For instance, deploying these tools for content moderation of online platforms or for other governance tasks raises normative questions about free-speech that requires the engagement from a diverse range of societal stakeholders and decision-makers. It is also crucial to consider the potential exploitation and abuse of these by malicious actors, such as authoritarian regimes, to limit free expression.  Mitigating this threat requires researchers and developers to be mindful of these considerations in the development of AI tools, actively engage with a diverse range of societal stakeholders in their development and deployment, and guard against their misuse by malign actors.

\bibliography{sample-ceur}

\begin{thebibliography}{48}
\expandafter\ifx\csname natexlab\endcsname\relax\def\natexlab#1{#1}\fi
\providecommand{\url}[1]{\texttt{#1}}
\providecommand{\href}[2]{#2}
\providecommand{\path}[1]{#1}
\providecommand{\DOIprefix}{doi:}
\providecommand{\ArXivprefix}{arXiv:}
\providecommand{\URLprefix}{URL: }
\providecommand{\Pubmedprefix}{pmid:}
\providecommand{\doi}[1]{\href{http://dx.doi.org/#1}{\path{#1}}}
\providecommand{\Pubmed}[1]{\href{pmid:#1}{\path{#1}}}
\providecommand{\bibinfo}[2]{#2}
\ifx\xfnm\relax \def\xfnm[#1]{\unskip,\space#1}\fi
\bibitem[{Scheufele(2014)}]{scheufele2014science}
\bibinfo{author}{D.~A. Scheufele},
\newblock \bibinfo{title}{Science communication as political communication},
\newblock \bibinfo{journal}{Proceedings of the National Academy of Sciences} \bibinfo{volume}{111} (\bibinfo{year}{2014}) \bibinfo{pages}{13585--13592}.
\bibitem[{Allgaier(2019)}]{allgaier2019science}
\bibinfo{author}{J.~Allgaier},
\newblock \bibinfo{title}{Science and environmental communication on youtube: Strategically distorted communications in online videos on climate change and climate engineering},
\newblock \bibinfo{journal}{Frontiers in communication} \bibinfo{volume}{4} (\bibinfo{year}{2019}) \bibinfo{pages}{446007}.
\bibitem[{Gounaridis and Newell(2024)}]{gounaridis2024social}
\bibinfo{author}{D.~Gounaridis}, \bibinfo{author}{J.~P. Newell},
\newblock \bibinfo{title}{The social anatomy of climate change denial in the united states},
\newblock \bibinfo{journal}{Scientific Reports} \bibinfo{volume}{14} (\bibinfo{year}{2024}) \bibinfo{pages}{2097}.
\bibitem[{IPCC(2022)}]{RN15}
\bibinfo{author}{IPCC}, \bibinfo{title}{Climate Change 2022: Impacts, Adaptation and Vulnerability}, Summary for Policymakers, \bibinfo{publisher}{Cambridge University Press}, \bibinfo{address}{Cambridge, UK and New York, USA}, \bibinfo{year}{2022}.
\bibitem[{Lewandowsky(2021)}]{lewandowsky2021climate}
\bibinfo{author}{S.~Lewandowsky},
\newblock \bibinfo{title}{Climate change disinformation and how to combat it},
\newblock \bibinfo{journal}{Annual Review of Public Health} \bibinfo{volume}{42} (\bibinfo{year}{2021}) \bibinfo{pages}{1--21}.
\bibitem[{Treen et~al.(2020)Treen, Williams, and O'Neill}]{treen2020online}
\bibinfo{author}{K.~M.~d. Treen}, \bibinfo{author}{H.~T. Williams}, \bibinfo{author}{S.~J. O'Neill},
\newblock \bibinfo{title}{Online misinformation about climate change},
\newblock \bibinfo{journal}{Wiley Interdisciplinary Reviews: Climate Change} \bibinfo{volume}{11} (\bibinfo{year}{2020}) \bibinfo{pages}{e665}.
\bibitem[{Fore et~al.(2024)Fore, Singh, Lee, Pandey, Anastasopoulos, and Stamoulis}]{fore2024unlearning}
\bibinfo{author}{M.~Fore}, \bibinfo{author}{S.~Singh}, \bibinfo{author}{C.~Lee}, \bibinfo{author}{A.~Pandey}, \bibinfo{author}{A.~Anastasopoulos}, \bibinfo{author}{D.~Stamoulis},
\newblock \bibinfo{title}{Unlearning climate misinformation in large language models},
\newblock \bibinfo{journal}{arXiv preprint arXiv:2405.19563}  (\bibinfo{year}{2024}).
\bibitem[{Zhang et~al.(2024)Zhang, Sharma, Du, and Liu}]{zhang2024toward}
\bibinfo{author}{Y.~Zhang}, \bibinfo{author}{K.~Sharma}, \bibinfo{author}{L.~Du}, \bibinfo{author}{Y.~Liu},
\newblock \bibinfo{title}{Toward mitigating misinformation and social media manipulation in llm era},
\newblock in: \bibinfo{booktitle}{Companion Proceedings of the ACM on Web Conference 2024}, \bibinfo{year}{2024}, pp. \bibinfo{pages}{1302--1305}.
\bibitem[{CAAD(2024)}]{climate2024underperforming}
\bibinfo{author}{CAAD}, \bibinfo{title}{Underperforming \& unprepared}, \bibinfo{year}{2024}. \bibinfo{note}{Climate Action Against Disinformation's report highlights how platforms have responded to the EU legislation for online safety so far.}
\bibitem[{Ellison and Hugh(2024)}]{ellison2024climate}
\bibinfo{author}{T.~Ellison}, \bibinfo{author}{B.~Hugh}, \bibinfo{title}{Climate security and misinformation: A baseline}, \bibinfo{year}{2024}.
\bibitem[{Kreps et~al.(2022)Kreps, McCain, and Brundage}]{kreps2022all}
\bibinfo{author}{S.~Kreps}, \bibinfo{author}{R.~M. McCain}, \bibinfo{author}{M.~Brundage},
\newblock \bibinfo{title}{All the news that’s fit to fabricate: Ai-generated text as a tool of media misinformation},
\newblock \bibinfo{journal}{Journal of experimental political science} \bibinfo{volume}{9} (\bibinfo{year}{2022}) \bibinfo{pages}{104--117}.
\bibitem[{Marlow et~al.(2020)Marlow, Miller, and Roberts}]{marlow2020twitter}
\bibinfo{author}{T.~Marlow}, \bibinfo{author}{S.~Miller}, \bibinfo{author}{J.~T. Roberts},
\newblock \bibinfo{title}{Twitter discourses on climate change: exploring topics and the presence of bots}  (\bibinfo{year}{2020}).
\bibitem[{CCDH(2021)}]{toxictenten2021}
\bibinfo{author}{CCDH}, \bibinfo{title}{The toxic ten: How 10 fringe publishers fuel 69\% of digital climate change denial}, \bibinfo{year}{2021}. \bibinfo{note}{Report}.
\bibitem[{Romero-Vicente(2023)}]{romero2023factsheet}
\bibinfo{author}{A.~Romero-Vicente}, \bibinfo{title}{Platforms' policies on climate change misinformation}, \bibinfo{year}{2023}. \bibinfo{note}{Factsheet}.
\bibitem[{Coan et~al.(2021)Coan, Boussalis, Cook, and Nanko}]{coan2021computer}
\bibinfo{author}{T.~G. Coan}, \bibinfo{author}{C.~Boussalis}, \bibinfo{author}{J.~Cook}, \bibinfo{author}{M.~O. Nanko},
\newblock \bibinfo{title}{Computer-assisted classification of contrarian claims about climate change},
\newblock \bibinfo{journal}{Scientific reports} \bibinfo{volume}{11} (\bibinfo{year}{2021}) \bibinfo{pages}{22320}.
\bibitem[{Vu et~al.(2023)Vu, Baines, and Nguyen}]{vu2023fact}
\bibinfo{author}{H.~T. Vu}, \bibinfo{author}{A.~Baines}, \bibinfo{author}{N.~Nguyen},
\newblock \bibinfo{title}{Fact-checking climate change: An analysis of claims and verification practices by fact-checkers in four countries},
\newblock \bibinfo{journal}{Journalism \& Mass Communication Quarterly} \bibinfo{volume}{100} (\bibinfo{year}{2023}) \bibinfo{pages}{286--307}.
\bibitem[{Leippold et~al.(2024)Leippold, Vaghefi, Stammbach, Muccione, Bingler, Ni, Colesanti-Senni, Wekhof, Schimanski, Gostlow et~al.}]{leippold2024automated}
\bibinfo{author}{M.~Leippold}, \bibinfo{author}{S.~A. Vaghefi}, \bibinfo{author}{D.~Stammbach}, \bibinfo{author}{V.~Muccione}, \bibinfo{author}{J.~Bingler}, \bibinfo{author}{J.~Ni}, \bibinfo{author}{C.~Colesanti-Senni}, \bibinfo{author}{T.~Wekhof}, \bibinfo{author}{T.~Schimanski}, \bibinfo{author}{G.~Gostlow}, et~al.,
\newblock \bibinfo{title}{Automated fact-checking of climate change claims with large language models},
\newblock \bibinfo{journal}{arXiv preprint arXiv:2401.12566}  (\bibinfo{year}{2024}).
\bibitem[{Stiff and Johansson(2022)}]{stiff2022detecting}
\bibinfo{author}{H.~Stiff}, \bibinfo{author}{F.~Johansson},
\newblock \bibinfo{title}{Detecting computer-generated disinformation},
\newblock \bibinfo{journal}{International Journal of Data Science and Analytics} \bibinfo{volume}{13} (\bibinfo{year}{2022}) \bibinfo{pages}{363--383}.
\bibitem[{Ni et~al.(2024)Ni, Xue, Yue, Deng, Shah, Jain, Neubig, and You}]{ni2024mixeval}
\bibinfo{author}{J.~Ni}, \bibinfo{author}{F.~Xue}, \bibinfo{author}{X.~Yue}, \bibinfo{author}{Y.~Deng}, \bibinfo{author}{M.~Shah}, \bibinfo{author}{K.~Jain}, \bibinfo{author}{G.~Neubig}, \bibinfo{author}{Y.~You},
\newblock \bibinfo{title}{Mixeval: Deriving wisdom of the crowd from llm benchmark mixtures},
\newblock \bibinfo{journal}{arXiv preprint arXiv:2406.06565}  (\bibinfo{year}{2024}).
\bibitem[{Thulke et~al.(2024)Thulke, Gao, Pelser, Brune, Jalota, Fok, Ramos, van Wyk, Nasir, Goldstein, Tragemann, Nguyen, Fowler, Stanco, Gabriel, Taylor, Moro, Tsymbalov, de~Waal, Matusov, Yaghi, Shihadah, Ney, Dugast, Dotan, and Erasmus}]{thulke2024climategpt}
\bibinfo{author}{D.~Thulke}, \bibinfo{author}{Y.~Gao}, \bibinfo{author}{P.~Pelser}, \bibinfo{author}{R.~Brune}, \bibinfo{author}{R.~Jalota}, \bibinfo{author}{F.~Fok}, \bibinfo{author}{M.~Ramos}, \bibinfo{author}{I.~van Wyk}, \bibinfo{author}{A.~Nasir}, \bibinfo{author}{H.~Goldstein}, \bibinfo{author}{T.~Tragemann}, \bibinfo{author}{K.~Nguyen}, \bibinfo{author}{A.~Fowler}, \bibinfo{author}{A.~Stanco}, \bibinfo{author}{J.~Gabriel}, \bibinfo{author}{J.~Taylor}, \bibinfo{author}{D.~Moro}, \bibinfo{author}{E.~Tsymbalov}, \bibinfo{author}{J.~de~Waal}, \bibinfo{author}{E.~Matusov}, \bibinfo{author}{M.~Yaghi}, \bibinfo{author}{M.~Shihadah}, \bibinfo{author}{H.~Ney}, \bibinfo{author}{C.~Dugast}, \bibinfo{author}{J.~Dotan}, \bibinfo{author}{D.~Erasmus},
\newblock \bibinfo{title}{Climategpt: Towards ai synthesizing interdisciplinary research on climate change}  (\bibinfo{year}{2024}). \href{http://arxiv.org/abs/2401.09646}{{\tt arXiv:2401.09646}}.
\bibitem[{Lacombe et~al.(2023)Lacombe, Wu, and Dilworth}]{lacombe2023climatex}
\bibinfo{author}{R.~Lacombe}, \bibinfo{author}{K.~Wu}, \bibinfo{author}{E.~Dilworth},
\newblock \bibinfo{title}{Climatex: Do llms accurately assess human expert confidence in climate statements?},
\newblock \bibinfo{journal}{arXiv preprint arXiv:2311.17107}  (\bibinfo{year}{2023}).
\bibitem[{Gehrmann et~al.(2023)Gehrmann, Clark, and Sellam}]{gehrmann2023repairing}
\bibinfo{author}{S.~Gehrmann}, \bibinfo{author}{E.~Clark}, \bibinfo{author}{T.~Sellam},
\newblock \bibinfo{title}{Repairing the cracked foundation: A survey of obstacles in evaluation practices for generated text},
\newblock \bibinfo{journal}{Journal of Artificial Intelligence Research} \bibinfo{volume}{77} (\bibinfo{year}{2023}) \bibinfo{pages}{103--166}.
\bibitem[{Xiao et~al.(2024)Xiao, Deng, Lam, Eslami, Kim, Lee, and Liao}]{xiao2024human}
\bibinfo{author}{Z.~Xiao}, \bibinfo{author}{W.~H. Deng}, \bibinfo{author}{M.~S. Lam}, \bibinfo{author}{M.~Eslami}, \bibinfo{author}{J.~Kim}, \bibinfo{author}{M.~Lee}, \bibinfo{author}{Q.~V. Liao},
\newblock \bibinfo{title}{Human-centered evaluation and auditing of language models},
\newblock in: \bibinfo{booktitle}{Extended Abstracts of the CHI Conference on Human Factors in Computing Systems}, \bibinfo{year}{2024}, pp. \bibinfo{pages}{1--6}.
\bibitem[{Yang and Menczer(2024)}]{Yang_2024}
\bibinfo{author}{K.~Yang}, \bibinfo{author}{F.~Menczer},
\newblock \bibinfo{title}{Anatomy of an ai-powered malicious social botnet},
\newblock \bibinfo{journal}{Journal of Quantitative Description: Digital Media} \bibinfo{volume}{4} (\bibinfo{year}{2024}). \URLprefix \url{http://dx.doi.org/10.51685/jqd.2024.icwsm.7}. \DOIprefix\doi{10.51685/jqd.2024.icwsm.7}.
\bibitem[{Ferrara et~al.(2020)Ferrara, Chang, Chen, Muric, and Patel}]{ferrara2020characterizing}
\bibinfo{author}{E.~Ferrara}, \bibinfo{author}{H.~Chang}, \bibinfo{author}{E.~Chen}, \bibinfo{author}{G.~Muric}, \bibinfo{author}{J.~Patel},
\newblock \bibinfo{title}{Characterizing social media manipulation in the 2020 us presidential election},
\newblock \bibinfo{journal}{First Monday}  (\bibinfo{year}{2020}).
\bibitem[{Akhtar et~al.(2023)Akhtar, Masood, Ikram, and Kanhere}]{akhtar2023false}
\bibinfo{author}{M.~M. Akhtar}, \bibinfo{author}{R.~Masood}, \bibinfo{author}{M.~Ikram}, \bibinfo{author}{S.~S. Kanhere},
\newblock \bibinfo{title}{False information, bots and malicious campaigns: Demystifying elements of social media manipulations},
\newblock \bibinfo{journal}{arXiv preprint arXiv:2308.12497}  (\bibinfo{year}{2023}).
\bibitem[{De~Angelis et~al.(2023)De~Angelis, Baglivo, Arzilli, Privitera, Ferragina, Tozzi, and Rizzo}]{de2023chatgpt}
\bibinfo{author}{L.~De~Angelis}, \bibinfo{author}{F.~Baglivo}, \bibinfo{author}{G.~Arzilli}, \bibinfo{author}{G.~P. Privitera}, \bibinfo{author}{P.~Ferragina}, \bibinfo{author}{A.~E. Tozzi}, \bibinfo{author}{C.~Rizzo},
\newblock \bibinfo{title}{Chatgpt and the rise of large language models: the new ai-driven infodemic threat in public health},
\newblock \bibinfo{journal}{Frontiers in Public Health} \bibinfo{volume}{11} (\bibinfo{year}{2023}) \bibinfo{pages}{1166120}.
\bibitem[{Chen and Shu(2023)}]{chen2023can}
\bibinfo{author}{C.~Chen}, \bibinfo{author}{K.~Shu},
\newblock \bibinfo{title}{Can llm-generated misinformation be detected?},
\newblock \bibinfo{journal}{arXiv preprint arXiv:2309.13788}  (\bibinfo{year}{2023}).
\bibitem[{Mullappilly et~al.(2023)Mullappilly, Shaker, Thawakar, Cholakkal, Anwer, Khan, and Khan}]{mullappilly2023arabic}
\bibinfo{author}{S.~S. Mullappilly}, \bibinfo{author}{A.~Shaker}, \bibinfo{author}{O.~Thawakar}, \bibinfo{author}{H.~Cholakkal}, \bibinfo{author}{R.~M. Anwer}, \bibinfo{author}{S.~Khan}, \bibinfo{author}{F.~S. Khan},
\newblock \bibinfo{title}{Arabic mini-climategpt: A climate change and sustainability tailored arabic llm},
\newblock \bibinfo{journal}{arXiv preprint arXiv:2312.09366}  (\bibinfo{year}{2023}).
\bibitem[{Vaid et~al.(2022)Vaid, Pant, and Shrivastava}]{vaid-etal-2022-towards}
\bibinfo{author}{R.~Vaid}, \bibinfo{author}{K.~Pant}, \bibinfo{author}{M.~Shrivastava},
\newblock \bibinfo{title}{Towards fine-grained classification of climate change related social media text},
\newblock in: \bibinfo{editor}{S.~Louvan}, \bibinfo{editor}{A.~Madotto}, \bibinfo{editor}{B.~Madureira} (Eds.), \bibinfo{booktitle}{Proceedings of the 60th Annual Meeting of the Association for Computational Linguistics: Student Research Workshop}, \bibinfo{publisher}{Association for Computational Linguistics}, \bibinfo{address}{Dublin, Ireland}, \bibinfo{year}{2022}, pp. \bibinfo{pages}{434--443}. \URLprefix \url{https://aclanthology.org/2022.acl-srw.35}. \DOIprefix\doi{10.18653/v1/2022.acl-srw.35}.
\bibitem[{Laud et~al.(2023)Laud, Spokoyny, Corringham, and Berg-Kirkpatrick}]{laud2023climabench}
\bibinfo{author}{T.~Laud}, \bibinfo{author}{D.~Spokoyny}, \bibinfo{author}{T.~Corringham}, \bibinfo{author}{T.~Berg-Kirkpatrick},
\newblock \bibinfo{title}{Climabench: A benchmark dataset for climate change text understanding in english},
\newblock \bibinfo{journal}{arXiv e-prints}  (\bibinfo{year}{2023}) \bibinfo{pages}{arXiv--2301}.
\bibitem[{Pirozelli et~al.(2023)Pirozelli, José, Silveira, Nakasato, Peres, Brandão, Costa, and Cozman}]{pirozelli2023benchmarks}
\bibinfo{author}{P.~Pirozelli}, \bibinfo{author}{M.~M. José}, \bibinfo{author}{I.~Silveira}, \bibinfo{author}{F.~Nakasato}, \bibinfo{author}{S.~M. Peres}, \bibinfo{author}{A.~A.~F. Brandão}, \bibinfo{author}{A.~H.~R. Costa}, \bibinfo{author}{F.~G. Cozman}, \bibinfo{title}{Benchmarks for pir\'a 2.0, a reading comprehension dataset about the ocean, the brazilian coast, and climate change}, \bibinfo{year}{2023}. \href{http://arxiv.org/abs/2309.10945}{{\tt arXiv:2309.10945}}.
\bibitem[{Liang et~al.(2022)Liang, Bommasani, Lee, Tsipras, Soylu, Yasunaga, Zhang, Narayanan, Wu, Kumar et~al.}]{liang2022holistic}
\bibinfo{author}{P.~Liang}, \bibinfo{author}{R.~Bommasani}, \bibinfo{author}{T.~Lee}, \bibinfo{author}{D.~Tsipras}, \bibinfo{author}{D.~Soylu}, \bibinfo{author}{M.~Yasunaga}, \bibinfo{author}{Y.~Zhang}, \bibinfo{author}{D.~Narayanan}, \bibinfo{author}{Y.~Wu}, \bibinfo{author}{A.~Kumar}, et~al.,
\newblock \bibinfo{title}{Holistic evaluation of language models},
\newblock \bibinfo{journal}{arXiv preprint arXiv:2211.09110}  (\bibinfo{year}{2022}).
\bibitem[{Stiennon et~al.(2020)Stiennon, Ouyang, Wu, Ziegler, Lowe, Voss, Radford, Amodei, and Christiano}]{stiennon2020learning}
\bibinfo{author}{N.~Stiennon}, \bibinfo{author}{L.~Ouyang}, \bibinfo{author}{J.~Wu}, \bibinfo{author}{D.~Ziegler}, \bibinfo{author}{R.~Lowe}, \bibinfo{author}{C.~Voss}, \bibinfo{author}{A.~Radford}, \bibinfo{author}{D.~Amodei}, \bibinfo{author}{P.~F. Christiano},
\newblock \bibinfo{title}{Learning to summarize with human feedback},
\newblock \bibinfo{journal}{Advances in Neural Information Processing Systems} \bibinfo{volume}{33} (\bibinfo{year}{2020}) \bibinfo{pages}{3008--3021}.
\bibitem[{Zhou and Xu(2020)}]{zhou2020learning}
\bibinfo{author}{W.~Zhou}, \bibinfo{author}{K.~Xu},
\newblock \bibinfo{title}{Learning to compare for better training and evaluation of open domain natural language generation models},
\newblock in: \bibinfo{booktitle}{Proceedings of the AAAI Conference on Artificial Intelligence}, volume~\bibinfo{volume}{34}, \bibinfo{year}{2020}, pp. \bibinfo{pages}{9717--9724}.
\bibitem[{Ouyang et~al.(2022)Ouyang, Wu, Jiang, Almeida, Wainwright, Mishkin, Zhang, Agarwal, Slama, Ray et~al.}]{ouyang2022training}
\bibinfo{author}{L.~Ouyang}, \bibinfo{author}{J.~Wu}, \bibinfo{author}{X.~Jiang}, \bibinfo{author}{D.~Almeida}, \bibinfo{author}{C.~Wainwright}, \bibinfo{author}{P.~Mishkin}, \bibinfo{author}{C.~Zhang}, \bibinfo{author}{S.~Agarwal}, \bibinfo{author}{K.~Slama}, \bibinfo{author}{A.~Ray}, et~al.,
\newblock \bibinfo{title}{Training language models to follow instructions with human feedback},
\newblock \bibinfo{journal}{Advances in neural information processing systems} \bibinfo{volume}{35} (\bibinfo{year}{2022}) \bibinfo{pages}{27730--27744}.
\bibitem[{Christiano et~al.(2017)Christiano, Leike, Brown, Martic, Legg, and Amodei}]{christiano2017deep}
\bibinfo{author}{P.~F. Christiano}, \bibinfo{author}{J.~Leike}, \bibinfo{author}{T.~Brown}, \bibinfo{author}{M.~Martic}, \bibinfo{author}{S.~Legg}, \bibinfo{author}{D.~Amodei},
\newblock \bibinfo{title}{Deep reinforcement learning from human preferences},
\newblock \bibinfo{journal}{Advances in neural information processing systems} \bibinfo{volume}{30} (\bibinfo{year}{2017}).
\bibitem[{MediaBiasFactCheck(2024)}]{mbfc}
\bibinfo{author}{MediaBiasFactCheck}, \bibinfo{title}{Mediabiasfactcheck}, \bibinfo{year}{2024}. \URLprefix \url{https://www.mediabiasfactcheck.com}, \bibinfo{note}{accessed: 2024-09-09}.
\bibitem[{NewsGuard(2024)}]{newsguard}
\bibinfo{author}{NewsGuard}, \bibinfo{title}{Newsguard}, \bibinfo{year}{2024}. \URLprefix \url{https://www.newsguardtech.com/solutions/newsguard/}, \bibinfo{note}{accessed: 2024-09-09}.
\bibitem[{Zhu and Tiwari(2023)}]{zhu2023climate}
\bibinfo{author}{H.~Zhu}, \bibinfo{author}{P.~Tiwari},
\newblock \bibinfo{title}{Climate change from large language models},
\newblock \bibinfo{journal}{arXiv preprint arXiv:2312.11985}  (\bibinfo{year}{2023}).
\bibitem[{Kraus et~al.(2023)Kraus, Bingler, Leippold, Schimanski, Senni, Stammbach, Vaghefi, and Webersinke}]{kraus2023enhancing}
\bibinfo{author}{M.~Kraus}, \bibinfo{author}{J.~A. Bingler}, \bibinfo{author}{M.~Leippold}, \bibinfo{author}{T.~Schimanski}, \bibinfo{author}{C.~C. Senni}, \bibinfo{author}{D.~Stammbach}, \bibinfo{author}{S.~A. Vaghefi}, \bibinfo{author}{N.~Webersinke},
\newblock \bibinfo{title}{Enhancing large language models with climate resources},
\newblock \bibinfo{journal}{arXiv preprint arXiv:2304.00116}  (\bibinfo{year}{2023}).
\bibitem[{Achiam et~al.(2023)Achiam, Adler, Agarwal, Ahmad, Akkaya, Aleman, Almeida, Altenschmidt, Altman, Anadkat et~al.}]{achiam2023gpt}
\bibinfo{author}{J.~Achiam}, \bibinfo{author}{S.~Adler}, \bibinfo{author}{S.~Agarwal}, \bibinfo{author}{L.~Ahmad}, \bibinfo{author}{I.~Akkaya}, \bibinfo{author}{F.~L. Aleman}, \bibinfo{author}{D.~Almeida}, \bibinfo{author}{J.~Altenschmidt}, \bibinfo{author}{S.~Altman}, \bibinfo{author}{S.~Anadkat}, et~al.,
\newblock \bibinfo{title}{Gpt-4 technical report},
\newblock \bibinfo{journal}{arXiv preprint arXiv:2303.08774}  (\bibinfo{year}{2023}).
\bibitem[{Dubey et~al.(2024)Dubey, Jauhri, Pandey, Kadian, Al-Dahle, Letman, Mathur, Schelten, Yang, Fan et~al.}]{dubey2024llama}
\bibinfo{author}{A.~Dubey}, \bibinfo{author}{A.~Jauhri}, \bibinfo{author}{A.~Pandey}, \bibinfo{author}{A.~Kadian}, \bibinfo{author}{A.~Al-Dahle}, \bibinfo{author}{A.~Letman}, \bibinfo{author}{A.~Mathur}, \bibinfo{author}{A.~Schelten}, \bibinfo{author}{A.~Yang}, \bibinfo{author}{A.~Fan}, et~al.,
\newblock \bibinfo{title}{The llama 3 herd of models},
\newblock \bibinfo{journal}{arXiv preprint arXiv:2407.21783}  (\bibinfo{year}{2024}).
\bibitem[{Rajpurkar et~al.(2018)Rajpurkar, Jia, and Liang}]{rajpurkar2018know}
\bibinfo{author}{P.~Rajpurkar}, \bibinfo{author}{R.~Jia}, \bibinfo{author}{P.~Liang},
\newblock \bibinfo{title}{Know what you don't know: Unanswerable questions for squad},
\newblock \bibinfo{journal}{arXiv preprint arXiv:1806.03822}  (\bibinfo{year}{2018}).
\bibitem[{Bavaresco et~al.(2024)Bavaresco, Bernardi, Bertolazzi, Elliott, Fern{\'a}ndez, Gatt, Ghaleb, Giulianelli, Hanna, Koller et~al.}]{bavaresco2024llms}
\bibinfo{author}{A.~Bavaresco}, \bibinfo{author}{R.~Bernardi}, \bibinfo{author}{L.~Bertolazzi}, \bibinfo{author}{D.~Elliott}, \bibinfo{author}{R.~Fern{\'a}ndez}, \bibinfo{author}{A.~Gatt}, \bibinfo{author}{E.~Ghaleb}, \bibinfo{author}{M.~Giulianelli}, \bibinfo{author}{M.~Hanna}, \bibinfo{author}{A.~Koller}, et~al.,
\newblock \bibinfo{title}{Llms instead of human judges? a large scale empirical study across 20 nlp evaluation tasks},
\newblock \bibinfo{journal}{arXiv preprint arXiv:2406.18403}  (\bibinfo{year}{2024}).
\bibitem[{Vidgen et~al.(2024)Vidgen, Agrawal, Ahmed, Akinwande, Al-Nuaimi, Alfaraj, Alhajjar, Aroyo, Bavalatti, Blili-Hamelin et~al.}]{vidgen2024introducing}
\bibinfo{author}{B.~Vidgen}, \bibinfo{author}{A.~Agrawal}, \bibinfo{author}{A.~M. Ahmed}, \bibinfo{author}{V.~Akinwande}, \bibinfo{author}{N.~Al-Nuaimi}, \bibinfo{author}{N.~Alfaraj}, \bibinfo{author}{E.~Alhajjar}, \bibinfo{author}{L.~Aroyo}, \bibinfo{author}{T.~Bavalatti}, \bibinfo{author}{B.~Blili-Hamelin}, et~al.,
\newblock \bibinfo{title}{Introducing v0. 5 of the ai safety benchmark from mlcommons},
\newblock \bibinfo{journal}{arXiv preprint arXiv:2404.12241}  (\bibinfo{year}{2024}).
\bibitem[{Matz et~al.(2024)Matz, Teeny, Vaid, Peters, Harari, and Cerf}]{matz2024potential}
\bibinfo{author}{S.~Matz}, \bibinfo{author}{J.~Teeny}, \bibinfo{author}{S.~S. Vaid}, \bibinfo{author}{H.~Peters}, \bibinfo{author}{G.~Harari}, \bibinfo{author}{M.~Cerf},
\newblock \bibinfo{title}{The potential of generative ai for personalized persuasion at scale},
\newblock \bibinfo{journal}{Scientific Reports} \bibinfo{volume}{14} (\bibinfo{year}{2024}) \bibinfo{pages}{4692}.
\bibitem[{Rojas et~al.(2024)Rojas, Algra-Maschio, Andrejevic, Coan, Cook, and Li}]{rojas2024augmented}
\bibinfo{author}{C.~Rojas}, \bibinfo{author}{F.~Algra-Maschio}, \bibinfo{author}{M.~Andrejevic}, \bibinfo{author}{T.~Coan}, \bibinfo{author}{J.~Cook}, \bibinfo{author}{Y.-F. Li},
\newblock \bibinfo{title}{Augmented cards: A machine learning approach to identifying triggers of climate change misinformation on twitter},
\newblock \bibinfo{journal}{arXiv preprint arXiv:2404.15673}  (\bibinfo{year}{2024}).

\end{thebibliography}

\newpage
\onecolumn
\appendix
\clearpage
\section{Appendix}

\subsection{Climate Change Keywords}\label{a.2}
A full list of the compiled climate change keywords identified by climate experts that are used to scrape and filter relevant climate change articles: climate change - climate crisis - climate effects - climate hoax - climate policy - climate resilience - climate science - climate summit - global warming - greenhouse gas - greenhouse gases - IPCC - green energy - climate hypocrisy - paris agreement - paris climate - net zero - net-zero - COP26 - climate conversation - climate test - climate gap - climate activists - climate activist - clean energy - climate negotiations - climate deal - green new deal - climate conference - green technology - green tech - climate fearmongering - climate fears - climate anxiety - carbon capture.

\subsection{CARDS Taxonomy of False or Misleading Claims}\label{a.1}

\begin{figure}[ht]
  \centering
  \includegraphics[width=\textwidth]{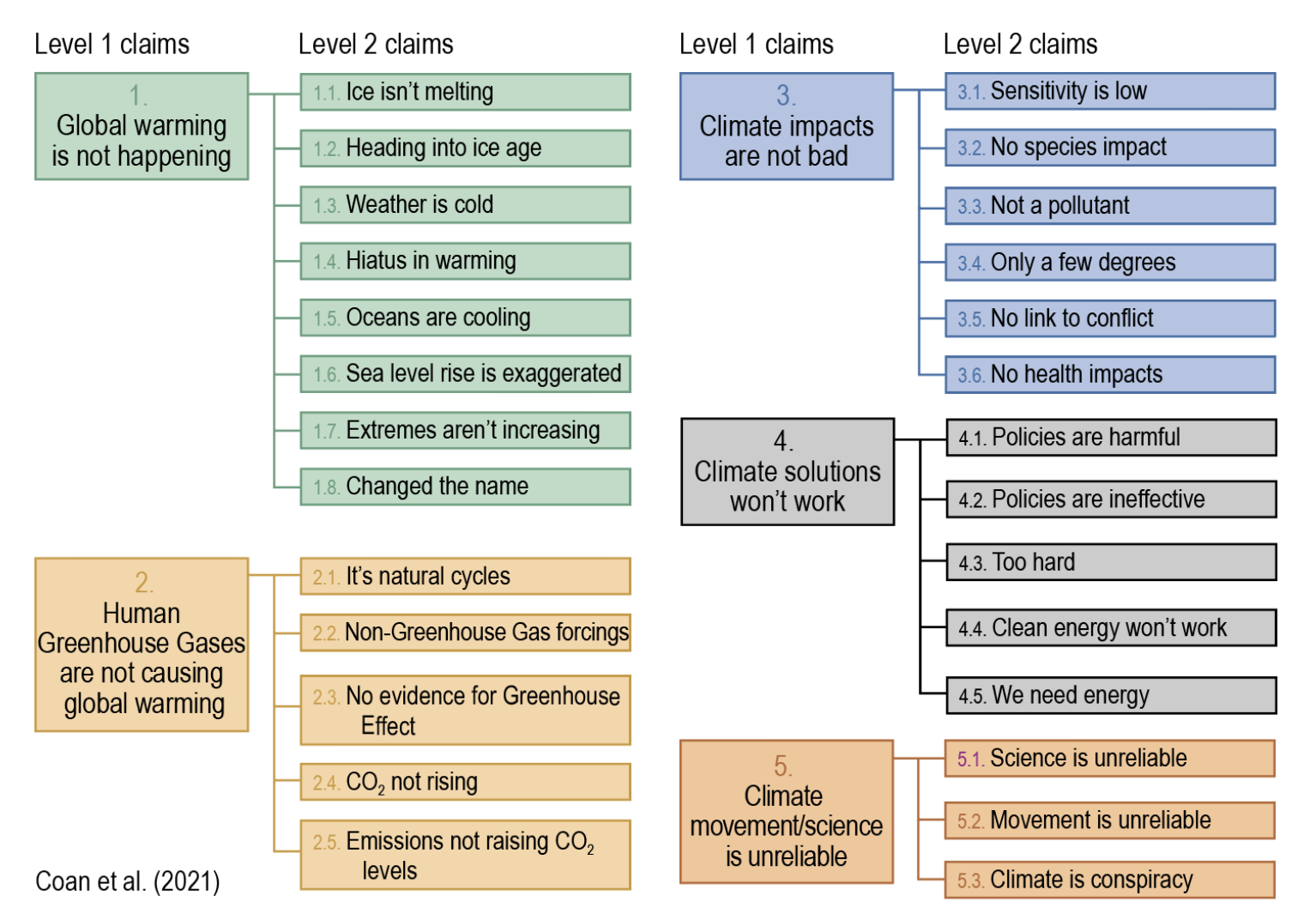}
  \captionsetup{width=\textwidth}
  \caption{Taxonomy of false or misleading claims published by \protect\citet{coan2021computer}}
  \label{fig:taxonomy-of-claims}
\end{figure}
\clearpage

\subsection{Prevalence of Domains in The Social Media Data}\label{a.3}
\begin{table}[htbp!]
\centering
\small
\captionsetup{width=\textwidth}
\begin{tabularx}{\textwidth}{l>{\centering\arraybackslash}X>{\centering\arraybackslash}X>{\centering\arraybackslash}l}

\toprule
\textbf{Domain} & \textbf{Number of Articles} & \textbf{\% Prevalence} & \textbf{Bias} \\
\midrule
newsbreak.com & 208,855 & 24.37\% & Questionable Source \\
freerepublic.com & 47,236 & 5.51\% & Right Bias \\
theepochtimes.com & 32,536 & 3.79\% & Questionable Source \\
foxnews.com & 29,598 & 3.45\% & Right Bias \\
beforeitsnews.com & 25,824 & 3.01\% & Questionable Source \\
breitbart.com & 25,001 & 2.91\% & Questionable Source \\
zerohedge.com & 21,853 & 2.55\% & Conspiracy-pseudocience \\
washingtonexaminer.com & 18,348 & 2.14\% & Right Bias \\
washingtontimes.com & 15,614 & 1.82\% & Questionable Source \\
patriotpost.us & 14,488 & 1.69\% & Right Bias \\
newsmax.com & 11,571 & 1.35\% & Questionable Source \\
americanthinker.com & 10,500 & 1.22\% & Questionable Source \\
wnd.com & 10,053 & 1.17\% & Questionable Sources \\
lawenforcementtoday.com & 9,291 & 1.08\% & Right Bias \\
shorenewsnetwork.com & 9,016 & 1.05\% & Right Bias \\
sott.net & 8,560 & 0.99\% & Conspiracy-pseudocience \\
dailycaller.com & 8,479 & 0.98\% & Right Bias \\
wattsupwiththat.com & 8,202 & 0.95\% & Conspiracy-pseudocience \\
bizpacreview.com & 8,028 & 0.93\% & Right Bias \\
townhall.com & 7,910 & 0.92\% & Questionable Source \\
lifesitenews.com & 7,562 & 0.88\% & Questionable Source \\
thelibertybeacon.com & 7,464 & 0.87\% & Conspiracy-pseudocience \\
noqreport.com & 7,294 & 0.85\% & Questionable source \\
dailywire.com & 5,446 & 0.63\% & Questionable Source \\
westernjournal.com & 5,406 & 0.63\% & Questionable Source \\
\bottomrule
\end{tabularx}
\caption{Top 25 low credible domains, their prevalence, and bias category accounting for 65.85\% of the total number of articles in the social media dataset described in Section \ref{3.2}.}
\label{tab:domain_data}
\end{table}
\clearpage

\subsection{System Prompt}\label{a.4}
\renewcommand{\lstlistingname}{Prompt}
\begin{lstlisting}[language=Python, caption={System prompt used for zero-shot classification of claims on the CARDS test dataset. The prompt is framed based on the coding manual retrieved from the supplimantary material of \protect\cite{coan2021computer}.}, label=list_gpt,  numbers=none, breaklines=true]
Overview:
---------
CARDS: Computer Assisted Recognition of Denial and Skepticism, is a machine learning project. Our aim is to train a computer to automatically detect and categorize misinformation about climate change. The end goal is that a computer can look at some text and successfully identify any climate misinformation - and even identify specific denialist claims. If successful, this will enable us to travel back in time and build a history of climate misinformation, including when myths originated and how they've evolved over time. It will also enable us to spot new publishing of denialist claims in real-time.

Context:
--------
Use the following coding rubric to answer the task assigned to you:
[
  {
    "code": "1_1",
    "identifier": 6,
    "claim": "Ice/permafrost/snow cover isn't melting"
  },
  {
    "code": "1_2",
    "identifier": 11,
    "claim": "We're heading into an ice age/global cooling"
  },
  {
    "code": "1_3",
    "identifier": 12,
    "claim": "Weather is cold/snowing"
  },
  {
    "code": "1_4",
    "identifier": 13,
    "claim": "Climate hasn't warmed/changed over the last (few) decade(s)"
  },
  {
    "code": "1_5",
    "identifier": 14,
    "claim": "Oceans are cooling/not warming"
  },
  {
    "code": "1_6",
    "identifier": 15,
    "claim": "Sea level rise is exaggerated/not accelerating"
  },
  {
    "code": "1_7",
    "identifier": 16,
    "claim": "Extreme weather isn't increasing/has happened before/isn't linked to climate change"
  },
  {
    "code": "1_8",
    "identifier": 17,
    "claim": "They changed the name from global warming' to climate change"
  },
  {
    "code": "2_1",
    "identifier": 18,
    "claim": "It's natural cycles/variation"
  },
  {
    "code": "2_2",
    "identifier": 24,
    "claim": "It's non-greenhouse gas human climate forcings (aerosols, land use)"
  },
  {
    "code": "2_3",
    "identifier": 25,
    "claim": "There's no evidence for greenhouse effect/carbon dioxide driving climate change"
  },
  {
    "code": "2_4",
    "identifier": 76,
    "claim": "C02 is not rising/ocean pH is not falling"
  },
  {
    "code": "2_5",
    "identifier": 78,
    "claim": "Human CO2 emissions are miniscule/not raising atmospheric CO2"
  },
  {
    "code": "3_1",
    "identifier": 31,
    "claim": "Climate sensitivity is low/negative feedbacks reduce warming"
  },
  {
    "code": "3_2",
    "identifier": 32,
    "claim": "Species/plants/reefs aren't showing climate impacts yet/are benefiting from climate"
  },
  {
    "code": "3_3",
    "identifier": 35,
    "claim": "C02 is beneficial/not a pollutant"
  },
  {
    "code": "3_4",
    "identifier": 37,
    "claim": "It's only a few degrees (or less)"
  },
  {
    "code": "3_5",
    "identifier": 38,
    "claim": "Climate change does not contribute to human conflict/threaten national security"
  },
  {
    "code": "3_6",
    "identifier": 39,
    "claim": "Climate change doesn't negatively impact health"
  },
  {
    "code": "4_1",
    "identifier": 40,
    "claim": "Climate policies (mitigation or adaptation) are harmful"
  },
  {
    "code": "4_2",
    "identifier": 46,
    "claim": "Climate policies are ineffective/flawed"
  },
  {
    "code": "4_3",
    "identifier": 53,
    "claim": "It's too hard to solve"
  },
  {
    "code": "4_4",
    "identifier": 55,
    "claim": "Clean energy technology/biofuels won't work"
  },
  {
    "code": "4_5",
    "identifier": 58,
    "claim": "People need energy (e.g., from fossil fuels/nuclear)"
  },
  {
    "code": "5_1",
    "identifier": 59,
    "claim": "Climate-related science is uncertain/unsound/unreliable (data, methods & models)"
  },
  {
    "code": "5_2",
    "identifier": 64,
    "claim": "Climate movement is alarmist/wrong/political/biased/hypocritical (people or groups)"
  }
]

Task:
-----
Classify whether a text excerpt belong to one of the claims outlined in the Context section provided 
in JSON format. Only respond in a JSON  format outlined below and don't make things up beyond what 
is given to you in the context. Below is the formatted JSON response template:
{
  "code": "CODE",
  "identifier": IDENTIFIER,
  "claim": "CLAIM"
}

If no claim is present in the text, just return a formatted json response like this one:
{
  "code": "0_0",
  "identifier": 0,
  "claim": "no claim"
}


This the end of the instructions. Now you will be provided a question with an excerpt of text and asked 
to identify the claim to which it belongs to.
"""
\end{lstlisting}
\newpage

\twocolumn
\subsection{User Prompt}\label{a.5}
\renewcommand{\lstlistingname}{Prompt}
\begin{lstlisting}[language=Python, caption={User prompt to classify paragraphs from articles into their corresponding claim label. The place holder \{text\} gets populated with the paragraph text at inference. }, label={list_gpt},  numbers=none, breaklines=true]
Question: To what claim does the following text belongs to?

{text}
    
Answer:
\end{lstlisting}

\begin{lstlisting}[language=Python, caption={Example user prompt illustrating how the paragraph is passed as part of the prompt }, label=list_gpt,  numbers=none, breaklines=true]
Question: To what claim does the following text belongs to?

What we are experiencing is outside of anything 
humans have seen on our planet and the only 
explanation that makes any real sense is that
it is due to human actions.
    
Answer:
\end{lstlisting}

\subsection{Fine-tuning Prompts}\label{a.6}
\subsubsection{System prompt}
\begin{lstlisting}[language=Python, caption={System prompt used as part of fine-tuning that describes for the model the task of classfying false or misleading claims about climate change.}, label=list_gpt,  numbers=none, breaklines=true]
You are an expert in classifying false and 
misleading claims about climate change in news media. You are asked to classify whether a text excerpt belongs to one of the following labels separated by a comma: 0_0, 1_1, 1_2, 1_3, 1_4,
1_5, 1_6, 1_7, 1_8, 2_1, 2_2, 2_3, 2_4, 2_5, 3_1, 
3_2, 3_3, 3_4, 3_5, 3_6, 4_1, 4_2, 4_3, 4_4, 
4_5, 5_1, 5_2.
Your answer must only include the classification 
label with no additional details

\end{lstlisting}

\subsubsection{Fine-tuning prompt structure}
\begin{lstlisting}[language=Python, caption={A template request that includes the system, user, and assistant messags that were used to fine-tune the model. All requests were sent in JSON format that include all three messages.}, label=list_gpt,  numbers=none, breaklines=true]
{
  "role": "system",
  "content": "You are an expert in classifying false and misleading claims about climate change in news media. You are asked to classify whether a text excerpt belongs to one of the following labels separated by a comma: 0_0, 1_1, 1_2, 1_3, 1_4, 1_5, 1_6, 1_7, 1_8, 2_1, 2_2, 2_3, 2_4, 2_5, 3_1, 3_2, 3_3, 3_4, 3_5, 3_6, 4_1, 4_2, 4_3, 4_4, 4_5, 5_1, 5_2. Your answer must only include the classification label with no additional details."
},
{
  "role": "user",
  "content": {text}
},
{
  "role": "assistant",
  "content": {claim}
}
\end{lstlisting}

\subsection{Inference Request with a Fine-Tuned Model}\label{a.7}
\begin{lstlisting}[language=Python, caption={A sample request sent to the fine-tuned GPT-3.5-turbo model to classify the paragraph in the user message to the corresponding claim label.}, label=list_gpt,  numbers=none, breaklines=true]
{
  "role": "system",
  "content": "You are an expert in classifying false and misleading claims about climate change in news media. You are asked to classify whether a text excerpt belongs to one of the following labels separated by a comma: 0_0, 1_1, 1_2, 1_3, 1_4, 1_5, 1_6, 1_7, 1_8, 2_1, 2_2, 2_3, 2_4, 2_5, 3_1, 3_2, 3_3, 3_4, 3_5, 3_6, 4_1, 4_2, 4_3, 4_4, 4_5, 5_1, 5_2. Your answer must only include the classification label with no additional details."
},
{
  "role": "user",
  "content": "What we are experiencing is outside of anything humans have seen on our planet and the only explanation that makes any real sense is that it is due to human actions"
},
{
  "role": "assistant",
  "content": ""
}
\end{lstlisting}

\subsection{Prompt Template for Open Source Models}\label{a.8}
\begin{lstlisting}[language=Python, caption={Prompt template used for open-source models in zero-shot classification task of false or misleading claims about climate change.}, label=list_gpt,  numbers=none, breaklines=true]
You are an expert in classifying false and 
misleading claims about climate change in news media. Classify the following text into one of the 27 classes.

Question: To which claim does the following text belongs to?

{text}

Classes:
0_0: no claim.
1_1: Ice/permafrost/snow cover isn't melting.
1_2: We're heading into an ice age/global cooling
1_3: Weather is cold/snowing.
1_4: Climate hasn't warmed/changed over the last (few) decade(s).
1_5: Oceans are cooling / not warming.
1_6: Sea level rise is exaggerated/not accelerating.
1_7: Extreme weather isn't increasing/has happened before/isn't linked to climate change.
1_8: They changed the name from global warming ' to climate change.
2_1: They changed the name from global warming' to climate change.
2_2: It 's non - greenhouse gas human climate forcings ( aerosols , land use ).
2_3: There's no evidence for greenhouse effect/carbon dioxide driving climate change.
2_4: C02 is not rising / ocean pH is not falling. 
2_5: Human CO2 emissions are miniscule / not raising atmospheric CO2.
3_1: Climate sensitivity is low/negative feedbacks reduce warming.
3_2: Species/plants/reefs aren't showing climate impacts yet/are benefiting from climate.
3_3: C02 is beneficial/not a pollutant.
3_4: It 's only a few degrees ( or less ).
3_5: Climate change does not contribute to human conflict / threaten national security.
3_6: Climate change doesn 't negatively impact health. 
4_1: Climate policies (mitigation or adaptation) are harmful
4_2: Climate policies are ineffective/flawed
4_3: It 's too hard to solve.
4_4: Clean energy technology/biofuels won't work
4_5: People need energy (e.g., from fossil fuels/nuclear)
5_1: Climate-related science is uncertain/unsound/unreliable (data, methods & models)
5_2: Climate movement is alarmist/wrong/political/biased/hypocritical (people or groups).

Respond only with a single class label. DO NOT 
add extra details.

Answer:
\end{lstlisting}

\subsection{Performance comparison between GPT-4o, CARDS, and Fine-tuned GPT-3.5-turbo}
\begin{table}[htbp!]
  \centering
  \begin{tabular}{lccc}
    \hline
    \textbf{Metric} & \textbf{GPT-3.5-turbo} & \textbf{GPT-4o} & \textbf{RoBERTa$_{\mathit{large}}$} \\
    \hline
    Precision & 0.88 & 0.70 & 0.82 \\
    Recall    & 0.81 & 0.84 & 0.75 \\
    F1-Score  & 0.84 & 0.75 & 0.77 \\
    \hline
  \end{tabular}
  \caption{Comparing the performance of GPT-3.5-turbo and GPT-4o in classifying false or misleading claims about climate change in paragraphs belonging to the CARDS test dataset across three classification metrics: precision, recall, and F1-Score. The results were also evaluated against the RoBERTa$_{\mathit{large}}$ CARDS model on the same test set. All evaluation metrics are macro averaged across the five categories of super-claims.}
  \label{tab:comparison}
\end{table}

\clearpage
\onecolumn
\subsection{Proportion of Valid LLM Responses by Model}
\begin{figure}[ht]
  \centering
  \includegraphics[width=\textwidth]{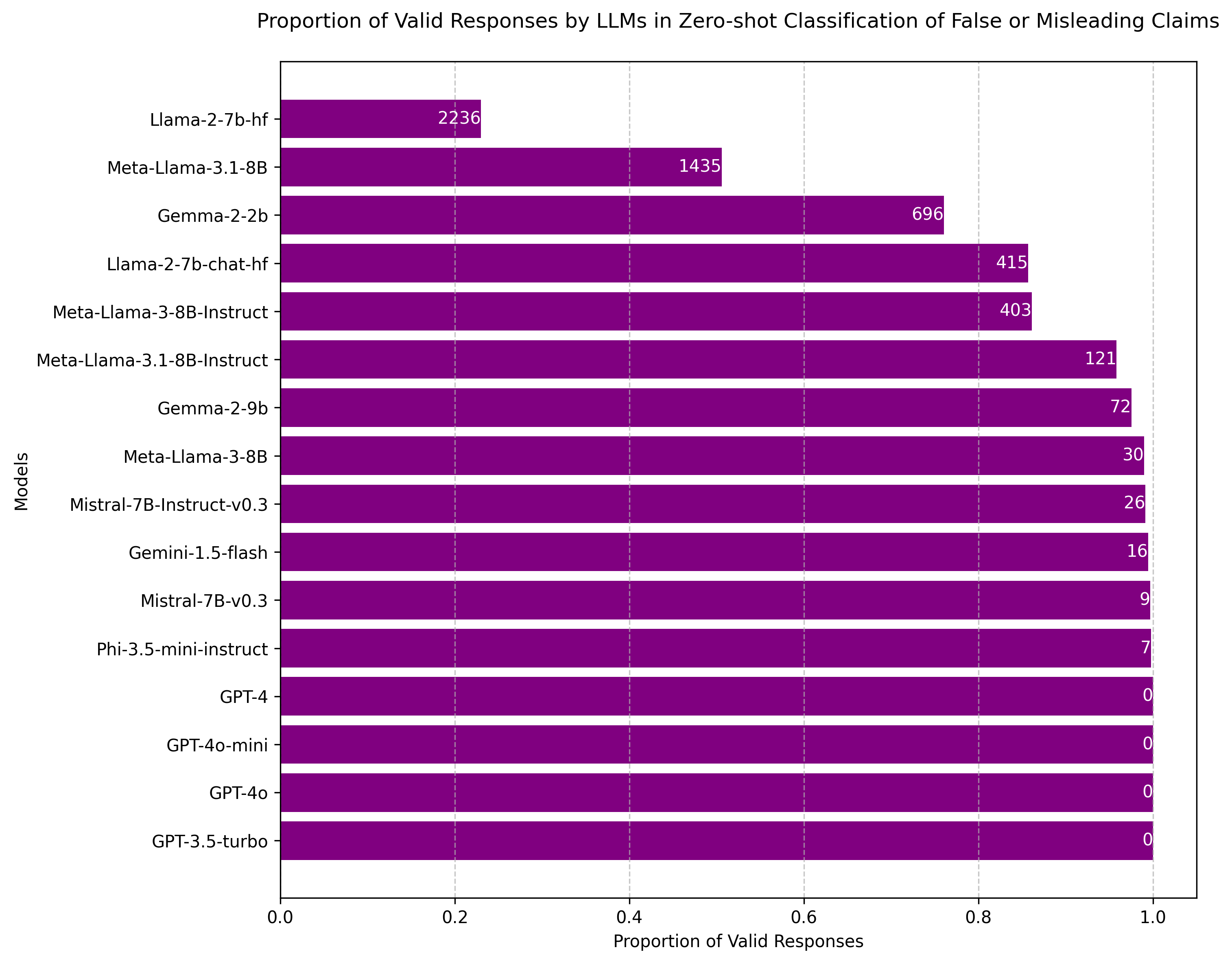}
  \captionsetup{width=\textwidth, justification=centering}
  \caption{Proportion of valid responses for each LLM in the zero-shot task described in section \ref{4.1} to classify false or misleading claims about climate change in the CARDS test dataset. Numbers inside the bar plot represent the number of invalid responses by each model.}
  \label{fig:taxonomy-of-claims}
\end{figure}


\clearpage
\subsection{Sample of Classified Claims from Social Media Data}
The table below shows a sample of claims sourced from social media data, as described in section \ref{4.3}. These claims have been annotated by two climate communication experts using the CARDS taxonomy of false or misleading claims about climate change, as proposed by \protect\cite{coan2021computer} (illustrated in Appendix \ref{a.2}). In addition, the table shows instances of the alignment (\includegraphics[scale=0.25]{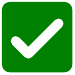}) and misalignment (\includegraphics[scale=0.15]{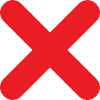}) of the GPT-3.5-turbo model -- fine-tuned on the expert-annotated dataset from \protect\cite{coan2021computer} -- and the CARDS model with the expert annotations.

\begingroup
\setlength\LTleft{0pt} 
\setlength\LTright{0pt} 
\begin{longtable}{m{4.5cm} m{4.5cm} m{3.5cm} >{\centering\arraybackslash}m{1.2cm} >{\centering\arraybackslash}m{1.2cm}}
  \toprule
  Paragraph & Domain & Experts Annotation & GPT-3.5-turbo (fine-tuned) & CARDS Model \\
  \midrule
  \endfirsthead
  \toprule
  Paragraph & Domain & Experts Annotation & GPT-3.5-turbo (fine-tuned) & CARDS Model \\
  \midrule
  \endhead
  \bottomrule
  \endfoot

  While there have been plenty of attempts by politicians and those in the media to both advance and avoid the recession narrative especially as election season approaches, evidence is mounting the next economic contraction is already underway. Does it matter whether experts politicians and political pundits agree to call the nation's current economic climate a recession is largely irrelevant to the average american. & \href{https://www.illinoispolicy.org/recession-depends-on-the-expert-but-its-bad-news-for-illinois/}{illinoispolicy.org} & No claim & \includegraphics[scale=0.25]{check_mark.pdf} & \includegraphics[scale=0.15]{red_cross.pdf} \\
  \\
  It seems pretty straightforward the supreme court is telling congress to do its job and pass laws just as the constitution requires it to do and thereby face the consequences of those laws. Political accountability is essential if laws are to be made after all. There is no accountability for bureaucrats who cannot be fired under civil service regulations. & \href{https://www.americanthinker.com/blog/2022/07/justice_kagans_sophomoric_misunderstanding_of_emwest_virginia_v_environmental_protection_agencyem.html}{americanthinker.com} & No claim & \includegraphics[scale=0.25]{check_mark.pdf} & \includegraphics[scale=0.25]{check_mark.pdf} \\
  \\
  Some extent as the penetration of intermittent resources increases. California policy makers have determined resource investment resource allocations and how and when grid improvements are made to enhance reliability. To blame extreme weather for causing the current concerns seems to be quite a reach. I suspect that a careful and fair examination of the weather data would should that the weather triggering such concerns this was not anything extraordinary considering historical weather patterns. & \href{https://wattsupwiththat.com/2022/09/13/will-california-learn-to-avoid-peak-rolling-blackouts/}{wattsupwiththat.com} & Global warming is not happening & \includegraphics[scale=0.25]{check_mark.pdf} & \includegraphics[scale=0.25]{check_mark.pdf} \\
  \\
  It is not a typo, more people die from cold temperatures than warm or hot temperatures. Contrary to the fear mongering assertions in the hill and time the overwhelming scientific evidence shows it is cold not heat that kills. Therefore, a modestly warmer world with shorter less severe winters should result in fewer premature deaths from disease viruses pandemics hunger and other natural causes. & \href{https://wattsupwiththat.com/2022/12/12/anthony-fauci-merges-covid-climate-infectious-diseases-largely-the-result-of-human-encroachment-on-nature-often-aided-by-climate-changes/}{wattsupwiththat.com} & Global warming is not happening & \includegraphics[scale=0.25]{check_mark.pdf} & \includegraphics[scale=0.15]{red_cross.pdf} \\
  \\
  Did summer just recently become hot? I seem to remember summers being hot as a kid. We didn't have air conditioning until I was almost out of the house so maybe the heat exhaustion affected my memory. My fuzzy memory seems to recall sweltering days and muggy nights though the climate alarmists would have us believe that hot summers are a relatively new phenomenon. We like to blame everything on the boogeyman known as climate change these days & \href{https://trendingpolitics.com/cbs-blames-childhood-obesity-onclimate-change-robm/}{trendingpolitics.com} & Humans greenhouse gases are not causing global warming & \includegraphics[scale=0.25]{check_mark.pdf} & \includegraphics[scale=0.25]{check_mark.pdf} \\
  \\
  ..hothouses are just fine with those levels.  When we breathe, we take in oxygen and exhale CO2. The concentration of CO2 in our exhaled breath is about 38,000ppm. That level would be a bit high for continued breathing, but if CO2 really were toxic, our breath would kill us. During the age of dinosaurs, an average level was about 900ppm, or over twice the concentration today.  In those times, the Earth was generally warmer than today – mostly tropical or semitropical as far as we can tell. & \href{https://www.conservativedailynews.com/2022/03/as-misinformation-reigns-now-is-the-climate-of-our-discontent/}{conservativedailynews.com} & Humans greenhouse gases are not causing global warming & \includegraphics[scale=0.25]{check_mark.pdf} & \includegraphics[scale=0.15]{red_cross.pdf} \\
  \\
  ..co2 also known as nature's fertilizer has produced a bounty of bumper crops. Australia reports record wheat barley and canola crops and near record sorghum crop. India the world's second largest producer of wheat expects record exports this year. Brazil expects record corn. Russia with another record crop will be the world's largest wheat exporter. Had things just been left alone free of the clutches of the globalists there would be more than enough food for everyone at affordable price & \href{https://www.americaoutloud.com/everything-bad-thats-happening-to-the-u-s-economy-is-part-of-the-globalists-plan/}{americaoutloud.com} & Climate impacts are not bad & \includegraphics[scale=0.25]{check_mark.pdf} & \includegraphics[scale=0.15]{red_cross.pdf} \\
  \\
  Carbon dioxide is a “nutrient to life on earth,” according to the Center for the Study of Carbon Dioxide and Global Change. Instead of chasing CO2 pipeline dreams, Tim Benson, policy analyst for The Heartland Institute, recommends ending subsidies and tax credits to carbon-capture technologies and removing “detrimental regulations” on reliable energy sources. He points out that “CO2 emissions in the United States have been relatively flat since 1990” while natural gas consumption has exploded.& \href{https://thenewamerican.com/why-should-you-care-about-carbon-capture/}{thenewamerican.com} & Climate impacts are not bad & \includegraphics[scale=0.25]{check_mark.pdf} & \includegraphics[scale=0.25]{check_mark.pdf} \\
  \\
  Advocating for an energy tax while soliciting massive government handouts for special interests is destructive, ineffective, and unaffordable. The Representatives Suggested They Will Be on the Outs with the Chamber of Commerce: “By pushing radical policy positions like a national energy tax, the Business Roundtable will quickly find itself alongside other fading organizations who lost their way.” Climate pricing is a globalist idea. It is a tax, but they call it climate pricing to obfuscate the very real effect it has on Americans.& \href{https://www.independentsentinel.com/?p=360964}{independentsentinel.com} & Climate solutions won’t work & \includegraphics[scale=0.25]{check_mark.pdf} & \includegraphics[scale=0.25]{check_mark.pdf} 
  \\
  \\
    ..unrestricted hunting of polar bears, is never mentioned in the media, Greenpeace, or politicians who say the polar bear is going extinct due to melting ice in the Arctic. In fact, the polar bear population has increased from 6,000 to 8,000 in 1973 to 30,000 to 50,000 today. This is not disputed," Moore said. Moore said that he does not pretend to know everything and predict the future with confidence like many in the "climate emergency" business claim they can do."I believe the human population has always been vulnerable to people who predict doom with false stories," &\href{https://www.sott.net/article/471890-Former-Greenpeace-founder-Patrick-Moore-says-climate-change-based-on-false-narratives}{sott.net} & Climate solutions won’t work & \includegraphics[scale=0.25]{check_mark.pdf} & \includegraphics[scale=0.15]{red_cross.pdf} \\
  \\
  Sky News host Chris Smith said, "While every fear-mongering greenie is saying we've never seen flooding like we have in recent years… the report found the opposite." He explained the politically correct conversations in recent years have stated without basis that "whatever disaster" is happening, "we've never seen anything like this before." & \href{https://www.wnd.com/2022/09/international-study-destroys-hoax-climate-emergency/}{wnd.com} & The climate movement and/or science are unreliable & \includegraphics[scale=0.25]{check_mark.pdf} & \includegraphics[scale=0.15]{red_cross.pdf} \\
  \\
  
  ..massive corporations and even single elite individuals who own private jets and fly around the planet to scold us about pollution. This is an extremely inconvenient truth for those who stand at their pulpits and demean the common folks for causing global warming — while they literally create more carbon in just hours than most people do in an entire year. Lest we forget, these are the same people telling us to eat bugs. & \href{https://beforeitsnews.com/health/2022/11/elite-fly-400-private-jets-to-cop27-sponsored-by-largest-plastic-polluter-in-world-to-lecture-you-about-climate-3047796.html}{beforeitsnews.com} & The climate movement and/or science are unreliable & \includegraphics[scale=0.25]{check_mark.pdf} & \includegraphics[scale=0.15]{red_cross.pdf} \\
\end{longtable}\label{model-expert-comparison-on-sm-data}
\endgroup


\end{document}